\documentclass[10pt,final,journal,letterpaper,twoside,twocolumn]{IEEEtran}
\usepackage{amsmath}
\usepackage{amssymb}
\usepackage[pdftex]{graphicx}
\usepackage{verbatim}
\usepackage{epsfig}
\usepackage{multirow}

\newtheorem{theorem}{Theorem}
\newtheorem{corollary}{Corollary}
\newtheorem{example}{Example}

\begin{document}
\title{The effect of fading, channel inversion, and threshold scheduling on ad hoc networks}
\author{Steven~Weber,~\IEEEmembership{Member,~IEEE,}
Jeffrey~G.~Andrews,~\IEEEmembership{Senior Member,~IEEE,}
Nihar~Jindal,~\IEEEmembership{Member,~IEEE}%
\thanks{Manuscript last saved: \today.}%
\thanks{Steven~Weber is with Drexel University.  Jeffrey~G.~Andrews is with The University of Texas at Austin. Nihar Jindal is with The University of Minnesota.}\thanks{Earlier versions of some of this material appeared in the 2006 Globecom and the 2006 Allerton conferences.}
\thanks{This work is supported under an NSF collaborative research grant awarded to the three authors (NSF grant \#0635003 (Weber), \#0634979 (Andrews), and  \#0634763 (Jindal)), and by the DARPA IT-MANET program, Grant W911NF-07-1-0028 (Andrews, Jindal, Weber).}}

%\markboth{Submitted: IEEE Transactions on Information Theory}{Weber,
%Andrews, and Jindal}

\maketitle

%%%%%%%%%%%%%%%%%%%%%%%%%%%%%%%%%%%%%%%%%%%%%%%%%%%%%%%%%%%%%%%%%%%%%%%
% ABSTRACT
%%%%%%%%%%%%%%%%%%%%%%%%%%%%%%%%%%%%%%%%%%%%%%%%%%%%%%%%%%%%%%%%%%%%%%%

\begin{abstract}
This paper addresses three issues in the field of ad hoc network capacity: the impact of $i)$ channel fading, $ii)$ channel inversion power control, and $iii)$ threshold--based scheduling on capacity.  Channel inversion and threshold scheduling may be viewed as simple ways to exploit channel state information (CSI) without requiring cooperation across transmitters.  We use the {\em transmission capacity} (TC) as our metric, defined as the maximum spatial intensity of successful simultaneous  transmissions subject to a constraint on the outage probability (OP).  By assuming the nodes are located on the infinite plane according to a Poisson process, we are able to employ tools from stochastic geometry to obtain asymptotically tight bounds on the distribution of the signal-to-interference (SIR) level, yielding in turn tight bounds on the OP (relative to a given SIR threshold) and the TC.  We demonstrate that in the absence of CSI, fading can significantly reduce the TC and somewhat surprisingly, channel inversion only makes matters worse. We develop a threshold-based transmission rule where transmitters are active only if the channel to their receiver is acceptably strong, obtain expressions for the optimal threshold, and show that this simple, fully distributed scheme can significantly reduce the effect of fading.
\end{abstract}

%\begin{IEEEkeywords}
%ad hoc networks, channel inversion, fading, threshold scheduling, transmission capacity
%\end{IEEEkeywords}

%%%%%%%%%%%%%%%%%%%%%%%%%%%%%%%%%%%%%%%%%%%%%%%%%%%%%%%%%%%%%%%%%%%%%%%
% 1. INTRODUCTION
%%%%%%%%%%%%%%%%%%%%%%%%%%%%%%%%%%%%%%%%%%%%%%%%%%%%%%%%%%%%%%%%%%%%%%%
\section{\label{sec:1} Introduction}

This paper addresses two issues of contemporary interest in the field of ad hoc network capacity. First, we characterize the effect of random channel variations, due both to shadowing/fading and to random distances between transmitter--receiver pairs.  Second, this paper considers the effect of local channel state information, namely through pairwise scheduling and power control. Through analysis we are able to obtain asymptotically tight lower and upper bounds on the transmission capacity. We anchor our discussion around three examples: lognormal shadowing, Rayleigh fading, and nearest neighbor transmissions (in a Poisson field).

Although fading without any channel state information (CSI) is shown to decrease capacity, fading might in fact enable an increase in capacity if it can be exploited. To investigate this we consider two simple ways to utilize local CSI: channel inversion power control and threshold--based scheduling. Both mechanisms require coordination only between each transmitter and its intended receiver, i.e., no coordination between transmitters is required.  Because the transmission capacity definition includes a universal SINR target, it may seem intuitive that channel inversion would be helpful, by saving power (and hence interference) from privileged links, and by providing assistance to underprivileged links to help them to avoid outage.  However, we prove that although channel inversion power control may help an individual link and does promote fairness, it lowers the network capacity as a whole.

Next, we characterize the potentially significant positive capacity impact of exploiting CSI for threshold--based scheduling.  In particular, each transmitter elects to transmit only if the channel to its receiver is acceptably strong.  Our results demonstrate that this simple scheduling rule provides significant capacity gains in a completely distributed manner. In effect, the threshold rule introduces {\em multi-user diversity} into the network by activating only those links with acceptable channel quality. A scientific contribution of this paper relative to prior work on ad hoc network scheduling is a novel framework for concisely and explicitly characterizing the effect of fading and scheduling in terms of the network and system parameters.

Some simplifying assumptions made in this paper are as follows.  First, we assume narrowband fading, i.e., each channel is affected by a single scalar gain.  Second, transmissions are slotted in time and multiple hop communication is not explicitly considered.  The goal of the considered framework is to quantify the maximum number of simultaneous successful transmissions per unit area; how these transmissions are used as far as routing packets over multiple hops is presently outside its scope.  Third, we ignore retransmissions, which will reduce the effective network capacity.  Finally, we assume that candidate transmitters are randomly located independent of one another, in particular according to a homogeneous Poisson point process.  The rest of our modeling assumptions are given in Section \ref{sec:3}.

%%%%%%%%%%%%%%%%%%%%%%%%%%%%%%%%%%%%%%%%%%%%%%%%%%%%%%%%%%%%%%%%%%%%%%%
% 1-A. Transmission and transport capacity
%%%%%%%%%%%%%%%%%%%%%%%%%%%%%%%%%%%%%%%%%%%%%%%%%%%%%%%%%%%%%%%%%%%%%%%
\subsection{\label{ssec:1a}Transmission capacity}

Throughout the paper we will employ {\em transmission capacity} (TC) as the primary performance metric.  The TC was introduced in \cite{WebYan2005}, and is defined as the maximum number of successful communication links that can be accommodated per unit area, subject to a specified constraint on the outage probability (OP) relative to a target signal to interference ratio (SIR)\footnote{Noise can also be included, but this is a negligible effect for interference-limited ad hoc networks, which is our case of interest.}. TC therefore quantifies the area spectral efficiency in an ad hoc network from an outage perspective.  A particular advantage of the TC framework is its amenability to precise analysis.  This allows the impact of physical layer effects (like fading) on link layer scheduling policies to be more precisely characterized.  Recently the TC has been employed to characterize capacity in a variety of scenarios, e.g., coverage \cite{VenHae06}, the capacity of irregular ad hoc networks \cite{GanHae06}, successive interference cancellation \cite{WebAnd2007}, or for better understanding of contention-based scheduling \cite{HasAnd07}.  In addition to ad hoc networks, the transmission capacity is also an appropriate metric for general open spectrum usage (e.g., Wi-Fi, cognitive radio) where many (non-cooperative) transmitter-receiver pairs operate in the same frequency band.

In \cite{WebYan2005}, the transmission capacity of an ad hoc network is studied for a network with path loss attenuation (no fading), fixed transmission power, and Aloha--style transmission attempts. In such a network the only source of randomness is the locations of the transmitters, modeled as a homogeneous Poisson process.  An outage occurs whenever the SINR falls below an SINR threshold $\beta$; in this simple setup the transmission capacity is
\begin{equation}
c(\epsilon) = \frac{\epsilon}{\pi \beta^{\frac{2}{\alpha}} d^{2}}  + \Theta(\epsilon^2),
\end{equation}
where $d$ is the fixed distance between each transmitter-receiver pair and $\alpha > 2$ is the path loss exponent.  Note that $c(\epsilon)$ has units of expected number of successful transmissions per unit area.

{\bf Relationship to transport capacity.}  The {\em transmission capacity} (TC) is closely related to the popular {\em transport capacity} metric introduced by Gupta and Kumar \cite{GupKum2000}.  The transport capacity is defined as the maximum weighted sum rate of communication over all pairs of $n$ nodes, where each pair's communication rate is weighted by the distance separating them.  A number of papers have studied transport capacity from an information theoretic perspective \cite{JovVis2004,XieKum2004,XueXie2005,XieKum2006,XueKum2006}, and the best result to date has shown that the transport capacity is $C_T(n) = \Theta(n)$ when nodes have a minimum distance separating them and the path loss exponent obeys $\alpha \geq 4$.  This minimum distance means that the area required for $n$ nodes is also $A(n) = \Theta(n)$.  As both transport capacity and the arena area are linear in $n$ it follows that $C_T(n)/A(n) = \Theta(1)$.  That is, the transport capacity per unit area is a constant, and has units of bit-meters/second per unit area.  The importance of this result is that $i)$ the transport capacity per unit area is independent of the number of nodes (for $n$ large), and $ii)$ local (one hop) communication is order optimal.

The transmission capacity can be converted into units of bit-meters/second per unit area by simply multiplying by the product of the average transmission rate times the average transmission distance.  In the outage setting considered here successful transmissions have rate $r = \log_2 (1 + \beta)$ (bps) and transmissions have a mean distance $d$ (meters). It follows that the transmission capacity is $c(\epsilon) r d$ in units of bit-meters/second per unit area.  We can write $c(\epsilon) r d = \Theta(1)$ to emphasize that the transmission capacity is order optimal, and thus order equivalent to the transport capacity\footnote{Recent work has shown that the $1/\sqrt{n}$ throughput scaling of multi-hop, which essentially corresponds to linear scaling of transport capacity in an extended network, can actually be exceeded for path loss exponents between 2 and 3 \cite{OzgLev2006sub}.  As a result, transmission capacity corresponds to an achievable rate that is not order--optimal for $2 \leq \alpha < 3$, but maximizing this quantity is still meaningful because multi-hop is currently the prevalent means of communication in ad hoc networks.}.  This constant depends upon the fundamental network parameters such as $\alpha, \beta, r, d, \epsilon$, as well as the particular technologies that are assumed, e.g. successive interference cancellation, CSI, power control, etc.

The contribution of the transport capacity framework is to prove optimality and achievability of $\Theta(1)$ bit-meters/second per unit area for as wide a class of networks as possible.  Because transport capacity seeks to make as few assumptions as possible regarding network behavior, the lower and upper constants obtained in proving the result are given only in terms of the path loss exponent and the minimum distance between nodes (see e.g., (8.1) in \cite{XueKum2006}).  Furthermore, the density of the network is not explicitly considered in works that have developed upper bounds to transport capacity scaling.  Our interest, on the other hand, is in determining the value of the unknown constant for various networks and transmission strategies (i.e., achievability schemes) of practical interest. The two metrics arise from distinct aims: transmission capacity aims to study the performance of a specific network (and gives performance expressions in terms of those specific model parameters), while transport capacity aims at establishing fundamental bounds over a broad class of networks.

%%%%%%%%%%%%%%%%%%%%%%%%%%%%%%%%%%%%%%%%%%%%%%%%%%%%%%%%%%%%%%%%%%%%%%%
% 1-B. Overview of main results
%%%%%%%%%%%%%%%%%%%%%%%%%%%%%%%%%%%%%%%%%%%%%%%%%%%%%%%%%%%%%%%%%%%%%%%
\subsection{\label{ssec:1b}Overview of main results}

\begin{figure}[ht]
\centering
\includegraphics[width=3.5in]{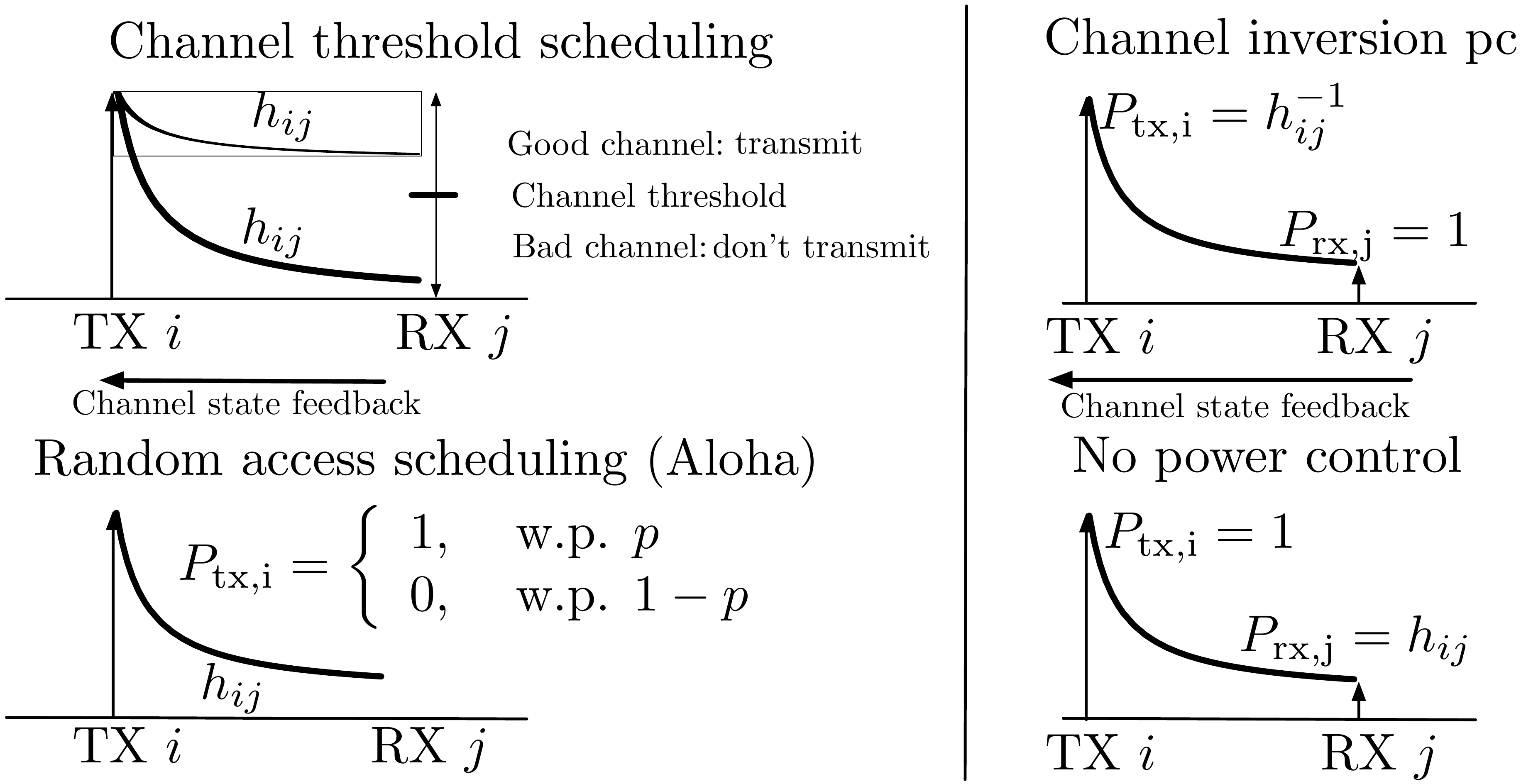}
\caption{Illustration of two uses of CSI to combat fading channels: threshold based scheduling (top left) and channel inversion power control (top right).   The bottom row gives the corresponding baseline mode (Aloha scheduling and fixed transmission power).  In channel threshold scheduling the transmitter elects to transmit provided the channel gain ($h_{ij}$) is above a specified threshold.  In channel inversion power control the transmitter selects a transmission power such that the received power is a specified value (here, $1$).}
\label{fig:0}
\end{figure}

\begin{table*}
\begin{center}
\caption{Mathematical summary of main results.}
\label{tab:0}
\begin{tabular}{|l|l|l|l|l|l|} \hline
{\bf Scheduling} & {\bf Power control} & {\bf Interference expression} & {\bf Outage prob. bound} & {\bf O.P. rate const.} & {\bf Notes} \\ \hline
Random access & No power control & $Y = \frac{1}{W_0} \sum_{i \in \Phi} \Psi_{i0} |X_i|^{-\alpha}$ & $q^l = 1 - \mathbb{E}[e^{-\Theta \mu(p)}]$ & $\Theta = \pi \mathbb{E}[\Psi^{\delta}]W^{-\delta} \beta^{\delta}$ & $\mu(p) = \lambda p$, $\delta = \frac{2}{\alpha}$ \\
Random access & Channel inversion & $Y = \sum_{i \in \Phi} \frac{1}{W_i} \Psi_{i0} |X_i|^{-\alpha}$ & $q^l = 1 - e^{-\theta \mu(p)}$ & $\theta = \mathbb{E}[\Theta]$ & \\
Threshold sched. & No power control & $Y = \frac{1}{W_{0|t}} \sum_{i \in \Phi} \Psi_{i0} |X_i|^{-\alpha}$ & $q^l = 1 - \mathbb{E}[e^{-\Theta(t) \mu(t)}]$ & $\Theta(t) = \pi \mathbb{E}[\Psi^{\delta}]W_t^{-\delta} \beta^{\delta}$ & $\mu(t) = \lambda \mathbb{P}(W > t)$   \\
Threshold sched. & Channel inversion & $Y = \sum_{i \in \Phi} \frac{1}{W_{i|t}} \Psi_{i0} |X_i|^{-\alpha}$ & $q^l = 1 - e^{-\theta(t) \mu(t)}$ & $\theta(t) = \mathbb{E}[\Theta(t)]$ & \\ \hline
\end{tabular}
\end{center}
\end{table*}

The main contribution of this paper is a comprehensive investigation of the effect of narrowband fading, both with and without CSI, on the transmission capacity of an ad hoc network. Two different strategies, channel inversion and threshold-scheduling, that potentially mitigate the effect of fading are considered, and all four combinations of the strategies are analyzed (see Figure \ref{fig:0}).

{\bf Summary of some of the mathematical results.}  In all four scenarios the received signal at a reference receiver $0$ at the origin is
\begin{equation}
\sqrt{P_0} Z_0 \sqrt{\Psi_{00}} D_0^{-\frac{\alpha}{2}} + \sum_{i \in \Phi} \sqrt{P_i} Z_i \sqrt{\Psi_{i0}} |X_i|^{-\frac{\alpha}{2}},
\end{equation}
where $D_0$ is the random distance separating the signal transmitter from the reference receiver, $\alpha$ is the path loss exponent, $Z_i$ is the signal intended for RX $i$, $P_i$ is the transmit power of TX $i$, $|X_i|$ is the distance from TX $i$ to RX 0, and $\Psi_{ij}$ is the fading coefficient on the link from TX $i$ to RX $j$. The corresponding SIR is given by:
\begin{equation}
{\rm SIR} = \frac{ P_0 \Psi_{00} d^{-\alpha}}{\sum_{i \in \Phi} P_i
\Psi_{i0} |X_i|^{-\alpha}}.
\end{equation}
We denote the received signal power at the reference receiver by $W_0$ with $W_0 = P_0 \Psi_{00} d^{-\alpha}$, and similarly use $W_i$ to denote the signal power at the $i$-th transmitter's receiver.  It is often convenient to work with the inverse of the SIR, i.e., the interference to signal ratio (ISR) , which we denote as $Y$. Using the definition of $W_0$, $Y$ can be expressed as:
\begin{equation}
Y = \frac{1}{W_0} \sum_{i \in \Phi} P_i \Psi_{i0} |X_i|^{-\alpha}.
\end{equation}
The probability of outage, $q$, is the probability the SIR falls below the SIR outage threshold $\beta$, or equivalently is the probability the ISR is too large: $q = \mathbb{P}(Y > y)$ for $y = 1/\beta$.

Table \ref{tab:0} summarizes some of the mathematical results for these four scenarios.  The first two columns identify the four scenarios of scheduling and power control.  The third column gives the expression for the random variable $Y$ denoting the ISR seen by a typical receiver at the origin.  The received signal power is unity for channel inversion, $W_0 = 1$.  Without power control the signal power is a random variable $W_0$ under random access, and a random variable $W_{0|t}$ under threshold scheduling.  The random variable $W_0$ is a random channel strength between a transmitter and its associated receiver; $W_{0|t}$ is the same but conditioned on the channel strength being above the threshold $t$.  The interference is summed over the interferers $\Phi$, which form a Poisson point process of intensity $\mu(p)$ (random access with probability $p$), or $\mu(t)$ (threshold scheduling with threshold $t$).  Without power control the individual interference contribution from interferer $i$ at location $X_i$ is simply the random channel gain $\Psi_{i0}$ times the pathloss $|X_i|^{-\alpha}$.  With power control the interference contribution is multiplied by the random variable $1/W_i$ (random access) or $1/W_{i|t}$ (threshold scheduling) representing the random power selected by node $i$ in compensating for the channel to $i$'s intended receiver.

The fourth column gives an explicit expression for an asymptotically tight lower bound on the outage probability, $q^l$.  The lower bounds for no power control involve the moment generating function (MGF) of a random variable $\Theta$ (for random access) or $\Theta(t)$ (for threshold scheduling), while the lower bounds for channel inversion are exponentially decreasing at rate $\theta$ (for random access) or $\theta(t)$ (for threshold scheduling).  We call $\Theta,\Theta(t),\theta,\theta(t)$ the rate constant for the outage probability decay (although $\Theta,\Theta(t)$ are random variables); the rate constants are given in the fifth column.  Finally, the sixth column gives the other expressions needed to translate the outage probability expressions back to fundamental model parameters.  First, $\lambda$ is the intensity of potential transmitters.  Under random access with transmission probability $p$ the intensity of actual transmitters is $\mu(p) = \lambda p$.  Under threshold scheduling with threshold $t$ the intensity of actual transmitters is $\mu(t) = \lambda \mathbb{P}(W > t)$, where $W$ is a random channel strength between a transmitter and its associated receiver.

{\bf Design implications of the mathematical results.}
The following list some of the design insights implied by the mathematical results.

\paragraph{Random access, no power control}  This is the baseline mode.  We compute the transmission capacity in this mode under fading channels and compare it with the transmission capacity under pure path loss.  The effect of fading is to reduce the transmission capacity by the factor
$\left(\mathbb{E}[\Psi^{\frac{2}{\alpha}}]\mathbb{E}[\Psi^{-\frac{2}{\alpha}}]\right)^{-1}$.  Fading of the desired signal has a {\em negative} effect while fading of interfering signals has a {\em positive} effect.  However, the net effect of fading is negative for any distribution because the above quantity is always less than unity. For example, in Rayleigh fading with $\alpha=3$ the loss is a factor of $0.41$.

\paragraph{Random access, channel inversion} Performing channel inversion actually decreases the transmission capacity relative to no power control.  One positive effect of channel inversion is that it assists with fairness.  If the distance between a transmitter-receiver pair is large compared to the average and/or the channel gain coefficient is small, the outage probability of this pair would be considerably higher than the network--wide average without channel inversion.  Channel inversion neutralizes distance and/or fading disadvantages and essentially puts all transmitter--receiver pairs on equal footing, but this fairness can come at the cost of reduced transmission capacity.  The capacity reduction is very small at low outage levels, but is much more significant at moderate and high outage levels.

\paragraph{Threshold scheduling, no power control} Threshold scheduling increases the transmission capacity relative to random access.  With threshold scheduling, users transmit only if the fading coefficient to the desired receiver is above some threshold $t$. Scheduling changes the distribution of $\Psi_{ii}$ (for all $i$) from the unconditional distribution $\Psi$ to the conditional distribution of $\Psi$ given $\Psi \geq t$ (but leaves the distribution of $\Psi_{ij}$ for $i \ne j$ unchanged).  Eliminating the fading coefficients below the threshold $t$ can significantly reduce outage for many fading distributions of interest (e.g., Rayleigh fading), and therefore can significantly increase the intensity of transmissions.  Performance with threshold-based scheduling can equal or even exceed that of a path-loss only network.

\paragraph{Threshold scheduling, channel inversion} Channel inversion in fact has little impact on the transmission capacity under threshold scheduling.  Threshold scheduling precludes transmission attempts by nodes in deep fades, and as such all transmitting nodes will require only moderate power to invert their channels.

The remainder of the paper is organized as follows.  Section \ref{sec:2} describes related work.  Section \ref{sec:3} introduces the mathematical model.  The TC for fading channels under randomized transmissions (with and without channel inversion) is derived in Section \ref{sec:4}; TC under threshold based transmission decisions (with and without channel inversion) is derived in Section \ref{sec:5}.  Section \ref{sec:6} contains the numerical and simulation results.  A brief conclusion is offered in Section \ref{sec:7}.  All proofs are found in the Appendix.

%%%%%%%%%%%%%%%%%%%%%%%%%%%%%%%%%%%%%%%%%%%%%%%%%%%%%%%%%%%%%%%%%%%%%%%
% 2. Related work
%%%%%%%%%%%%%%%%%%%%%%%%%%%%%%%%%%%%%%%%%%%%%%%%%%%%%%%%%%%%%%%%%%%%%%%
\section{\label{sec:2} Related work and Preliminaries}

%%%%%%%%%%%%%%%%%%%%%%%%%%%%%%%%%%%%%%%%%%%%%%%%%%%%%%%%%%%%%%%%%%%%%%%
% 2-B. Fading channels
%%%%%%%%%%%%%%%%%%%%%%%%%%%%%%%%%%%%%%%%%%%%%%%%%%%%%%%%%%%%%%%%%%%%%%%
\subsection{\label{ssec:2b}Fading channels}

Computing the TC under the assumed channel fading model involves computing the tail probability of the random SIR seen by a typical receiver.  The SIR can be viewed as the spatial analog of the familiar temporal power-law shot noise process, where the cumulative effect of the impulse response of Poisson driven shocks in time is replaced with the cumulative effect of the channel response of a Poisson driven set of interferers in space. Previous results on spatial shot noise processes in wireless networks have characterized the aggregate co-channel interference under distance attenuation with random fading as a stable random process \cite{SouSil1990,IloHat1998,BacBla2006}.  In \cite{BacBla2006}, an exact expression for the outage capacity in a Rayleigh fading environment, assuming randomized transmissions and no power control, is derived using the moment generating function of the interference power.  Interestingly, the lower bound to outage probability for the case of channel inversion in a Rayleigh fading environment exactly matches the expression in \cite{BacBla2006}; this is discussed in detail in Section \ref{ssec:4c}.

Our characterization of the TC under general fading models relies upon results from three distinct but related fields of study: stable distributions, shot noise processes, and spatial co-channel interference models.

{\bf Stable distributions.}
Stable distributions, introduced by L\'{e}vy in 1925 \cite{Lev1925}, are defined as distributions that are closed under convolution.   More precisely, the random variable $X$ is said to be stable if, for $X_1,X_2$ independent and identically distributed (iid) copies of $X$, there exist constants $a,b,c,d$ such that
\begin{equation}
\label{eqn:a}
a X_1 + b X_2 \stackrel{d}{=} c X + d,
\end{equation}
where the equality holds in distribution, see e.g., Shao and Nikias \cite{ShaNik1993}.  Except in special cases (e.g., Gaussian and Cauchy), there is no closed form expression for the PDF or CDF of a stable random variable.  Instead, the family is parameterized by its characteristic function.  For the sub-family of {\em symmetric} stable random variables (the case of relevance to us) the characteristic function is
\begin{equation}
\label{eqn:b}
\phi(t) = \mathbb{E} \left[ e^{i t X} \right] = \exp \left\{ - \gamma |t|^{\delta} \right\},
\end{equation}
where $\gamma > 0$ is dispersion parameter and $0 \leq \delta \leq 2$ is the characteristic or stability exponent.  Stable random variables with $\delta < 2$ have fractional moments given by
\begin{equation}
\label{eqn:c}
\mathbb{E}[|X|^p] \left\{ \begin{array}{ll} < \infty, \; & 0 \leq p \leq \delta, \\ = \infty, \; & p > \delta \end{array} \right. ,
\end{equation}
and $\mathbb{E}[|X|^p] < \infty$ for all $p \geq 0$ for the Gaussian case of $\delta = 2$ \cite{ShaNik1993}.  In particular all stable random variables (except the limiting Gaussian case) have infinite variance.  The importance of stable distributions is illuminated by the so-called {\em generalized CLT}: for $\{X_i\}$ iid and $\{a_n\},\{b_n\}$ with $a_n \to \infty$, then
\begin{equation}
\frac{1}{a_n} \sum_{i=1}^n X_i - b_n \stackrel{\mathcal{D}}{\to} X,
\end{equation}
iff $X$ is stable, where the convergence is in distribution \cite{ShaNik1993}.  Petropulu {\em et al.} \cite{PetPes2000} have further developed the implications of stable distributions on signal processing in communications.

{\bf Shot noise process.}  The shot noise process was first described by Schottky \cite{Sch1918} in 1918, and was soon applied to noise modeling in a wide variety of fields.  The general shot noise process, using the notation of Lowen and Teich \cite{LowTei1990}, is a functional
\begin{equation}
\label{eqn:d}
I(t) = \sum_{j = -\infty}^{\infty} h(t-t_j),
\end{equation}
where $\{t_j\}$ is a stationary Poisson process on $\mathbb{R}$ and $h(t)$ is the (linear, time-invariant) impulse response function.  Thus $I(t)$ is the superposition of responses seen at time $t$ caused by all previous times $t_j \leq t$.  Extensive work was done by Rice {\em et al.} from the 1940s through the 1970s to characterize the CDF and PDF of the random variable $I(t)$, e.g., \cite{Ric1944}.  More recent algorithms for computation are found in Gubner \cite{Gub1996}.  A characterization of the stochastic process $\{I(t), t \in \mathbb{R}\}$ is provided by Lowen and Teich \cite{LowTei1990} for the important case when $h(t)$ is a power law, i.e.,
\begin{equation}
\label{eqn:e}
h(t) = K t^{-\beta}, ~ 0 \leq A \leq t \leq B < \infty, ~ 0 < \beta \leq 2,
\end{equation}
and $K$ can be either deterministic or random.  They make the important observation that $I(t)$ is a stable random variable for certain values of $\beta,A,B$.  Their framework is restricted to the time-dimension, i.e., the points $t_j$ are times in a Poisson process on $\mathbb{R}^1$.

{\bf Spatial co-channel interference models.}  The use of spatial models for co-channel interference in packet radio (ad hoc) networks goes back at least to 1978 where Musa and Wasylkiwskyj \cite{MusWas1978} consider the impact of node locations on the aggregate interference.  This idea was further developed by Sousa and Silvester in a series of papers in the early 1990s \cite{SouSil1990,Sou1990,Sou1992}.  Sousa and Silvester characterize the aggregate co-channel interference as a stable distribution, although they do not mention anywhere that it is a shot-noise process.  Sousa's work is the first, as far as we are aware, to connect the aggregate interference generated by a distance dependent power law path loss channel model with a stable distribution (although spatial connections were made as early as 1919 by Holtsmark in astronomy \cite{Hol1919}, see \cite{ShaNik1993}). Ilow and Hatzinakos \cite{IloHat1998} characterize the impact of random channel effects on the aggregate co-channel interference.  They study the individual and combined impacts of lognormal shadowing and Rayleigh fading on the aggregate interference, where the interference effects are subject to a distance dependent path loss attenuation.  Our work extends theirs in that their focus was on identifying the impact of the fading model on the parameters of the characteristic function of the interference, while our focus is on link layer capacity and the benefit of CSI.  Baccelli {\em et al.} consider the impact of co-channel interference on link layer scheduling through the use of stochastic geometry \cite{BacBla2006}.  Their proposed multiple hop spatial reuse Aloha protocol maximizes a performance metric they call the spatial density of progress.  Their focus is on optimizing the power and access probability of Aloha protocols, whereas our focus is on characterizing the benefit of threshold scheduling with CSI on capacity.

%%%%%%%%%%%%%%%%%%%%%%%%%%%%%%%%%%%%%%%%%%%%%%%%%%%%%%%%%%%%%%%%%%%%%%%
% 2-C. Threshold scheduling with channel state information
%%%%%%%%%%%%%%%%%%%%%%%%%%%%%%%%%%%%%%%%%%%%%%%%%%%%%%%%%%%%%%%%%%%%%%%
\subsection{\label{ssec:2c}Threshold scheduling with channel state information}

Distributed channel-aware wireless scheduling has received extensive attention in the literature.  Much of this work is {\em game theoretic} in that transmission decisions of neighboring transmitters are coupled: an active neighboring interferer reduces the SIR seen by a receiver, which makes it less likely for that receiver's transmitter to transmit \cite{FelHub2006}.  The coupling of these decisions severely limits analytical tractability, and in practice can also result in adverse behavior and/or require considerable overhead.

In contrast, our approach precludes the transmitter interaction presumed in the game-theoretic approaches, i.e., transmission decisions are independent for each transmitter.  The success or failure of an individual transmission attempt, however, is of course dependent upon the joint decisions of all transmitters. In particular, we consider the realistic scenario where each user monitors the channel to just its desired recipient (either through channel reciprocity or a very low rate feedback channel), and then transmits opportunistically only when the channel strength is above a threshold.  We characterize the optimum threshold, and show that this simple approach increases the capacity significantly over a channel-blind Aloha approach.  The proposed threshold scheduling scheme is fully distributed and extremely simple, and can be viewed as an optimal scheduling approach under the specified side information constraint. Although the proposed approach is obviously suboptimal compared to a centralized scheduler with global channel knowledge, our scheme has the benefits of being more practical as well as yielding to analysis.  In particular, through stochastic geometry we obtain tight upper and lower bounds on the OP and TC under an arbitrary threshold, and from here obtain the TC-optimal threshold.

Prior work on quantifying ad hoc network capacity with transmitter CSI includes Toumpis and Goldsmith \cite{TouGol2003}, who determined that fading actually increases the achievable rate regions (as opposed to the overall ad hoc network capacity) by providing statistical diversity, since the best set of transmit-receive pairs can be selected. This however, would require a global centralized search which is impractical.  Toumpis and Goldsmith argue in a second paper that although fading reduced a transport capacity lower bound by a logarithmic factor, fading actually increased the overall network capacity \cite{TouGol2004}.  Using the transport capacity framework, some interesting recent results by Gowaiker {\em et al.} include a study on entirely random channels (no geometric dependence) that showed that shadowing or obstructions could increase the transport capacity \cite{GowHocSub}.  Xue and Xie \cite{XueXie2005} and Xie and Kumar \cite{XieKum2004} study fading channels with geometric considerations valid for path-loss exponents greater than three that supported their previous results in the absence of fading.  A recent review of this research thrust is found in the monograph by Georgiadis, Neely, and Tassiulas \cite{GeoNee2006}.  Essentially, in order to fully exploit fading, some delay must be introduced, which results in a delay-capacity tradeoff.  We will not consider this tradeoff in this paper, however.

%%%%%%%%%%%%%%%%%%%%%%%%%%%%%%%%%%%%%%%%%%%%%%%%%%%%%%%%%%%%%%%%%%%%%%%
% 3. Mathematical model
%%%%%%%%%%%%%%%%%%%%%%%%%%%%%%%%%%%%%%%%%%%%%%%%%%%%%%%%%%%%%%%%%%%%%%%
\section{\label{sec:3} Mathematical Model}

For a random variable $X$ we will write $F_X(x)$ for the cumulative distribution function (CDF), $f_X(x)$ for the probability density function (PDF), and $\bar{F}_X(x) = 1- F_X(x)$ for the complementary CDF (CCDF).  The exception to this rule is that $Q(z) = \mathbb{P}(Z > z)$ and $Q^{-1}(p)$ are used to denote the CCDF and inverse CCDF for $Z$ a standard normal $\mathcal{N}(0,1)$ random variable.  We write $X \sim F_X$ to denote that $X$ is a random variable with distribution $F_X$.  The superscripts $l,u$ will denote lower and upper bounds.  Table \ref{tab:1} summarizes the notation used throughout the paper.

\begin{table}
\caption{Summary of notation.}
\begin{center}
\begin{tabular}{|l|l|} \hline
Symbol & Meaning \\ \hline
$\alpha > 2$ & path loss exponent \\
$\delta =  \frac{2}{\alpha} < 1$ & stability exponent \\
$\Psi,\psi$ & (random, fixed) channel fading gain \\
$D,d$ & (random, fixed) transmitter to receiver distance \\
$\Pi = \{X_i\}$ & PPP of potential transmitters \\
$\lambda$ & intensity of $\Pi$ \\
$\Phi \subset \Pi$ & PPP of actual transmitters \\
$\mu \leq \lambda$ & intensity of $\Phi$ \\
$Y = 1/{\rm SIR}$ & normalized aggregate interference at origin \\
$\beta$ & minimum SIR required for successful reception \\
$y = 1/\beta$ & maximum normalized interference for reception \\
$q(\mu)$ & outage probability for a typical receiver \\
$\tau(\mu)$ & spatial throughput of successful transmissions \\
$c(\epsilon)$ & transmission capacity with outage constraint $\epsilon$ \\
$p = \mu/\lambda$ & transmission probability without CSI \\
$\kappa$ & constant governing performance without CSI \\
$\theta = \kappa \beta^{\delta}$ & constant governing performance without CSI\\
$W = \Psi D^{-\delta}$ & signal strength at receiver \\
$t$ & min. signal strength for transmission with CSI \\
$\mu(t) = \lambda \bar{F}_W(t)$ & intensity of transmissions with CSI \\
$\kappa(t)$ & function governing performance with CSI \\
$\theta(t) = \kappa(t) \beta^{\delta}$ & function governing performance with CSI \\
$\gamma(t) = \theta(t) \mu(t)$ & normalized intensity of transmissions with CSI \\
$\gamma^{-1}(g)$ & inverse of $\gamma(t)$ \\ \hline
\end{tabular}
\end{center}
\label{tab:1}
\end{table}

%%%%%%%%%%%%%%%%%%%%%%%%%%%%%%%%%%%%%%%%%%%%%%%%%%%%%%%%%%%%%%%%%%%%%%%
% 3-A. Channel model
%%%%%%%%%%%%%%%%%%%%%%%%%%%%%%%%%%%%%%%%%%%%%%%%%%%%%%%%%%%%%%%%%%%%%%%
\subsection{\label{ssec:3a} Channel model}

We consider a general class of channel models consisting of a deterministic distance-dependent path loss component with path loss exponent $\alpha > 2$, and a random distance-independent component.  In particular, let
\begin{equation}
\label{eqn:f}
h(d,\psi) = \psi d^{-\alpha}
\end{equation}
be the far-field attenuation in signal power over a distance $d$ with a channel gain $\psi$.  The distance independent channel gain, $\psi$, is assumed to be independent across channels and independent of the node position.  Note that this model has a singularity as $d \rightarrow 0$; this matter is discussed in some detail in \cite{SouSil1990} and \cite{BacBla2006}.  Because we consider the network from an outage perspective, such a singularity has only a negligible effect on our results.  For example, if an interferer is very close to a receiver, the above channel model would lead to an artificially small SIR.  However, the receiver would very likely be in outage even if the singularity was removed, and thus there is no effect on the outage probability.  Furthermore, we assume the distribution on transmitter--receiver pair separation distances precludes the possibility of nearby transmitter--receiver pairs.  Although the singularity at the origin is not physically meaningful, it turns out that retaining the singularity significantly simplifies the analysis without materially affecting the numerical and simulation results.  As explained further in the numerical and simulation results section (Section \ref{sec:6}), for purposes of analysis we will retain the singularity ($d_{\rm min} = 0$), but all our simulation results will employ $d_{\rm min} > 0$.  Our results will illustrate that the results are essentially unaffected by the singularity.

\begin{comment}
The step function $\mathbf{1}_{d > d_{\rm min}}$ is employed to prevent the physically anomalous singularity at the origin.  One can think of $d_{\rm min}$ as separating the near and far fields, so that $d_{\rm min}$ is on the order of several wavelengths.  The choice to set $h(d,m) = 0$ for $d < d_{\rm min}$ is rather arbitrary and inessential; it is also possible to select $h(d,m) = k$ for $d < d_{\rm min}$ (for some constant $k$), or to remove the singularity through the use of a function like $h(d,m) = \frac{m}{1+d^{\alpha}}$.  This matter is also discussed in \cite{SouSil1990} and \cite{BacBla2006}.  One can think of setting $d = 0$ for $d < d_{\rm min}$ as effectively cancelling the interference contribution of any interferer within distance $d_{\rm min}$ of a receiver.  When the probability of such an interferer actually being located so close is small it follows that the impact of the cancellation on performance will be minimal.
\end{comment}

For simplicity and analytical tractability we ignore background thermal noise. In an interference limited network the noise contribution is minimal.  Our earlier work \cite{WebYan2005} contained models with additive noise, and it was shown there was no appreciable effect unless the network was extremely sparse.  Of course, it is straightforward to numerically verify this claim.

We study network performance both with and without channel inversion.  In the absence of channel inversion we assume that unit power is employed; this results in no loss of generality because in the absence of additive noise increasing the power linearly increases both the signal and interference, leaving the SIR unaffected.  Under channel inversion each transmitter employs a power $p = 1/w$ where $w = h(\psi,d)$ is the channel gain connecting the transmitter with its intended receiver; this results in unit signal power at the intended receiver.  The impact of channel inversion on link layer performance for Poisson distributed transmitters is also addressed by Baccelli {\em et al.} \cite{BacBla2006}.

%%%%%%%%%%%%%%%%%%%%%%%%%%%%%%%%%%%%%%%%%%%%%%%%%%%%%%%%%%%%%%%%%%%%%%%
% 3-B. Network model
%%%%%%%%%%%%%%%%%%%%%%%%%%%%%%%%%%%%%%%%%%%%%%%%%%%%%%%%%%%%%%%%%%%%%%%
\subsection{\label{ssec:3b} Network model}

Consider a large ad hoc network, where the locations of {\em potential} transmitters at a typical point in time form a stationary Poisson point process (PPP) $\Pi = \{X_i\}$ on the plane $\mathbb{R}^2$.  The spatial density of the point process is denoted by $\lambda$, giving the average number of potential transmitters per unit area.  We also assume that each potential transmitter, $i$, has an associated intended receiver (not in $\Pi$), and we let the index $i$ refer to the pair consisting of transmitter $i$ and its associate receiver $i$.   The assumption that each potential transmitter has a receiver that is not a potential transmitter precludes the possibility of collisions where a transmitter attempts to communicate with another node that is already transmitting.

Let $\Psi_{ij}$ denote the random channel gain for the channel between the transmitter of pair $i$ and the receiver of pair $j$.  The channel gains are independent across both receivers ($\Psi_{ij}$ is independent of $\Psi_{ik}$), and across transmitters ($\Psi_{ji}$ is independent of $\Psi_{ki}$).  Let $F_{\Psi}$ be the common distribution for the channel gains.  Let $D_i$ represent the distance between the transmitter and intended receiver of pair $i$; the distances $\{D_i\}$ are assumed to be iid with common distribution $F_D$.  As discussed in the introduction, we restrict our attention to transmission policies where each transmitter's decision is made independent of the other transmitter decisions.  It follows that the relevant state information for each transmitter $i$'s decision is the pair $(\Psi_{ii},D_i)$ describing the channel with its intended receiver.

Our attention will focus on a (typical) reference receiver, without loss of generality assumed to be located at the origin, $o$.  The reference receiver and its associated transmitter are pair number $0$.  It follows that the performance will depend upon not only each pair's channel information (dictating which transmitters will elect to transmit), but also upon the channel information connecting each transmitter with the reference receiver at the origin (dictating the typical receiver performance).  We encode all this state information by forming the {\em marked} Poisson point process (MPPP)
\begin{equation}
\label{eqn:g}
\Pi = \{(X_i,D_i,\Psi_{ii},\Psi_{i0}), ~ i \in \mathbb{N}\}.
\end{equation}
Let $|X_i|$ denote the distance from each transmitter $i$ to the reference receiver at the origin.

The PPP $\Phi \subset \Pi$ denotes the set of actual interferers at the typical time under consideration.  Because the transmission decisions are made independently across transmitters and independent of their locations, it follows that $\Phi$ is also a stationary MPPP, albeit with a smaller intensity, denoted as $\mu \leq \lambda$.  We discuss transmission decision rules for obtaining $\Phi$ from $\Pi$ in Sections \ref{sec:4} (using random transmission decisions) and \ref{sec:5} (using threshold based transmission decisions).  Rather than work with the SIR we will instead work with its inverse, $Y = \frac{1}{{\rm SIR}_0}$, which can be thought of as the aggregate co-channel interference power normalized by the signal power.  The normalized aggregate interference seen at the reference receiver is
\begin{equation}
\label{eqn:h}
Y = \frac{\sum_{i \in \Phi} P_i h(|X|_i,\Psi_{i0})}{P_0 h(D_0,\Psi_{00})} = \frac{\sum_{i \in \Phi} P_i \Psi_{i0} |X_i|^{-\alpha}}{P_0 \Psi_{00} D_0^{-\alpha}},
\end{equation}
where $\{P_i\}$ are the transmission powers employed.  The SIR seen at the reference receiver is therefore
\begin{equation}
{\rm SIR}_0 = \frac{1}{Y} = \frac{P_0 \Psi_{00} D_0^{-\alpha}} {\sum_{i \in \Phi} P_i \Psi_{i0} |X_i|^{-\alpha}}.
\end{equation}

%%%%%%%%%%%%%%%%%%%%%%%%%%%%%%%%%%%%%%%%%%%%%%%%%%%%%%%%%%%%%%%%%%%%%%%
% 3-C. Performance metrics
%%%%%%%%%%%%%%%%%%%%%%%%%%%%%%%%%%%%%%%%%%%%%%%%%%%%%%%%%%%%%%%%%%%%%%%
\subsection{\label{ssec:3c}Performance metrics}

Three performance metrics are studied in this paper: the outage probability, the spatial throughput, and the transmission capacity.

{\bf Outage probability.}  A reception is assumed successful provided the SIR seen at the receiver exceeds a specified $\beta > 0$, with an outage resulting if this condition is not satisfied.  Let $q(\mu)$ denote the probability of outage when the intensity of attempted transmissions is $\mu$:
\begin{equation}
\label{eqn:i}
q(\mu) = \mathbb{P}({\rm SIR}_0 < \beta) = \mathbb{P}(Y > 1/\beta) = \bar{F}_Y(y),
\end{equation}
where $y = 1/\beta$ is the ISR requirement.

The SIR-based outage probability introduced above corresponds very simply to achievability in the information theoretic sense.  If all nodes are assumed to transmit Gaussian symbols and the channel is narrowband, the mutual information between the transmitting ($X_i$) and receiving node ($\hat{X}_i$) is given by:
\begin{equation}
\label{eqn:j}
I(X_i;\hat{X}_i | \Phi) = \log_2 (1 + {\rm SIR}_i),
\end{equation}
where ${\rm SIR}_i$ is the SIR seen by receiver $i$.  Since only the term $I(X_i;\hat{X}_i | \Phi)$ is considered, an implicit assumption is that multi-user interference is treated as noise (interference can be cancelled, see \cite{WebAnd2007}). Mutual information, or rate, is measured conditioned on channel conditions, node locations, the instantaneous set of transmitters, and the fading coefficients.  Thus, the quantity in (\ref{eqn:j}) measures the rate of reliable information flow from $X_i$ to $\hat{X}_i$ at a snapshot of the network.  Of course, this mutual information expression is only meaningful if the conditioning variables are fixed during transmission.  Most importantly, this requires that the time scale of fading be larger than packet durations.

In the outage formulation, the instantaneous mutual information is treated as a random variable (a function of random interferer locations and channel conditions) and an outage occurs whenever this random variable falls below the desired rate of communication.  Thus, for rate $R$ the outage probability is given by $P_{\rm out} = \mathbb{P}(I(X_i;\hat{X}_i | \Phi) < R)$.  Since there is a one-to-one mapping between mutual information and SIR in this expression, outage can equivalently be stated in terms of SIR, as in (\ref{eqn:i}) with $\beta = 2^R - 1$.

{\bf Spatial throughput.}  The spatial throughput is the expected spatial density of successful transmissions:
\begin{equation}
\label{eqn:k}
\tau(\mu) = \mu (1-q(\mu)),
\end{equation}
i.e., the product of the attempted transmission intensity ($\mu$) times the average probability of success ($1-q(\mu)$).

{\bf Transmission capacity.}  The spatial throughput often obscures the fact that high throughput is sometimes obtained at the expense of unacceptably high outage.  This is especially important in ad hoc networks as wasted transmissions both cause unnecessary interference for other nodes and they waste precious energy.  As a simple example of high throughput achieved through high outage, note that classic slotted Aloha has a throughput of the form $G e^{-G}$, which is maximized for an attempt rate of $G = 1$.  The optimal  throughput at $G = 1$ is $1/e \approx 0.32$, but the outage probability is $1 - 1/e \approx 0.68$.  Thus $68\%$ of all attempted transmissions must fail to achieve the optimal throughput.  For many important network applications, e.g., streaming media, high levels of outage are unacceptable, and as such it is desirable that the network operate in a low outage regime.  With this in mind, we define the {\em optimal contention density}, $\nu(\epsilon)$, as the maximum spatial density of {\em attempted} transmissions such that the corresponding outage probability is $\epsilon \in [0,1]$.  The parameter $\epsilon$ serves as a proxy for network quality of service.  The optimal contention density is found by solving $q(\nu) = \epsilon$ for $\nu$, i.e., $\nu = q^{-1}(\epsilon)$, where $q^{-1}$ is the inverse of (\ref{eqn:i}).  Having found the optimal contention density, we define the {\em transmission capacity} as the corresponding spatial density of {\em successful} transmissions,
\begin{equation}
\label{eqn:l}
c(\epsilon) = \nu(\epsilon)(1-\epsilon) = q^{-1}(\epsilon)(1-\epsilon).
\end{equation}
The advantage of the transmission capacity framework is that it yields the maximum throughput that can be obtained subject to a maximum permissible outage probability, i.e., a QoS requirement.

%%%%%%%%%%%%%%%%%%%%%%%%%%%%%%%%%%%%%%%%%%%%%%%%%%%%%%%%%%%%%%%%%%%%%%%
% 4. Performance without threshold scheduling
%%%%%%%%%%%%%%%%%%%%%%%%%%%%%%%%%%%%%%%%%%%%%%%%%%%%%%%%%%%%%%%%%%%%%%%
\section{\label{sec:4}Performance without threshold scheduling}

In this section we present analytical results for the performance metrics introduced in Section \ref{ssec:3c} when transmission decisions are made randomly; performance results with threshold scheduling decisions are given in Section \ref{sec:5}.  Under randomized transmissions the set of actual transmitters, $\Phi$, is obtained from the set of possible transmitters, $\Pi$, by each node electing to transmit at random with probability $p = \frac{\mu}{\lambda}$, for any desired $\mu \leq \lambda$.  We provide analytical results for performance with fixed (unit) power (Section \ref{ssec:4a}) and with channel inversion (Section \ref{ssec:4b}), and then provide detailed discussion (Section \ref{ssec:4d}) as well as examples (Section \ref{ssec:4c}).

%%%%%%%%%%%%%%%%%%%%%%%%%%%%%%%%%%%%%%%%%%%%%%%%%%%%%%%%%%%%%%%%%%%%%%%
% 4-A. Performance without threshold scheduling and without channel inversion
%%%%%%%%%%%%%%%%%%%%%%%%%%%%%%%%%%%%%%%%%%%%%%%%%%%%%%%%%%%%%%%%%%%%%%%
\subsection{\label{ssec:4a}Performance without threshold scheduling and without channel inversion}

Looking at the three performance metrics in Section \ref{ssec:3c} it is apparent that they each depend upon the distribution of $Y$ in (\ref{eqn:h}).  In the absence of channel inversion, the normalized aggregate interference seen by the reference receiver is
\begin{equation}
\label{eqn:m}
Y = \frac{1}{W_0} \sum_{i \in \Phi} \Psi_{i0} |X_i|^{-\alpha}, ~ W_0 = \Psi_{00} D_0^{-\alpha},
\end{equation}
where $W_0$ is the received signal power. Because transmission decisions are made by each node at random (independent of the channel state), it follows that each node electing to transmit has $W_0 \sim F_W$, where $F_W(w) = \mathbb{P}(W \leq w) = \mathbb{P}(\Psi D^{-\alpha} \leq w)$ is expressible in terms of the known distributions $F_{\Psi}$ and $F_D$.  The distribution of $Y$ may be expressed in terms of the distribution of $Y$ conditioned on $W_0$:
\begin{equation}
\label{eqn:n}
\bar{F}_Y(y) = \int_0^{\infty} \bar{F}_{Y|W}(y|w) {\rm d}F_W(w).
\end{equation}

Previous work by Ilow and Hatzinakos \cite{IloHat1998} has characterized the conditional distribution $\bar{F}_{Y|W}(y|w)$ as a stable distribution.  This forms the starting point of our analysis.  For easy reference we combine results from Theorems 1, 2, and 3 from Ilow and Hatzinakos \cite{IloHat1998} and repeat them below in a single theorem using our notation.

\begin{theorem}{\em (Ilow and Hatzinakos \cite{IloHat1998}).}
\label{thm:1}
{\em Under randomized transmissions and lacking channel inversion, the conditional distribution $\bar{F}_{Y|W}(y|w)$ in (\ref{eqn:n}) is symmetric stable with characteristic function given by (\ref{eqn:b}), with stability parameter $\delta = 2/\alpha < 1$ and dispersion parameter
\begin{equation}
\label{eqn:o}
\gamma(w) = \frac{\Gamma(2-\delta) }{1-\delta} \cos \left(\pi \frac{\delta}{2}\right) \kappa(w) \mu,
\end{equation}
for $\Gamma(\cdot)$ the Gamma function and
\begin{equation}
\label{eqn:p}
\kappa(w) = \pi \mathbb{E}[\Psi^{\delta}] w^{-\delta}.
\end{equation}}
\end{theorem}

As mentioned in the introduction, stable distributions are awkward to work with as they do not have closed form expressions for their PDF or CDF.  This motivates the importance of the bounds on the CCDF given in the next theorem.

\begin{theorem}
\label{thm:2}
{\em Under randomized transmissions and lacking channel inversion, the expressions $\bar{F}_Y^u,\bar{F}_Y^l$ are upper and lower bounds on the CCDF $\bar{F}_Y(y)$ of the random variable $Y$ in (\ref{eqn:m}):
\begin{eqnarray}
\bar{F}_Y^u(y) &=& 1 - \mathbb{E}\left[\left(1 - \frac{\frac{\delta}{2-\delta} K \mu y^{-\delta}} {(1 - \frac{\delta}{1-\delta} K \mu y^{-\delta})^2}\right)^+ e^{-K \mu y^{-\delta}}\right], \nonumber \\
\bar{F}_Y^l(y) &=& 1-\mathbb{E}\left[ e^{-K \mu y^{-\delta}}\right], \label{eqn:q}
\end{eqnarray}
where random variable $K$ is defined as:
\begin{equation}
\label{eqn:r}
K = \pi \mathbb{E}[\Psi^{\delta}]W^{-\delta} = \pi \mathbb{E}[\Psi^{\delta}]\Psi^{-\delta}D^2,
\end{equation}
$\Psi \sim F_{\Psi}$ and $D \sim F_D$.  The lower bound is asymptotically tight as $y \to \infty$ and the upper bound has an asymptotic
bounded error.  Specifically:
\begin{eqnarray}
\bar{F}_Y^l(y) &=& \kappa \mu y^{-\delta} + O(y^{-2 \delta}),
\label{eqn:s_lower} \\
\bar{F}_Y(y) &=& \kappa \mu y^{-\delta} + O(y^{-2 \delta}), \label{eqn:s_actual} \\
\bar{F}_Y^u(y) &=& \frac{2}{2-\delta} \kappa \mu y^{-\delta} + O(y^{-2\delta}), \label{eqn:s}
\end{eqnarray}
for
\begin{equation}
\label{eqn:t}
\kappa = \mathbb{E}[K] = \pi \mathbb{E}[\Psi^{\delta}] \mathbb{E}[\Psi^{-\delta}] \mathbb{E}[D^2].
\end{equation}
}
\end{theorem}

The full proof is provided in the Appendix. The lower bound $\bar{F}_Y^l(y)$ is the probability that a single term in the sum in (\ref{eqn:m}) is larger than $y$, i.e., the probability that there exists at least one {\em dominant} interferer that individually contributes enough interference to cause outage relative to threshold $y$.  Note that due to fading, a dominant interferer need not correspond to the nearest interferer.  Indeed, considering only the contribution of the nearest interferer gives a weaker bound. The upper bound $\bar{F}_Y^u(y)$ is obtained by application of the Chebychev inequality.  We now make several remarks on the theorem:

{\bf Asymptotic impact of channel variations.}  The impact of the random channel fading gains, $\{\Psi_{ij}\}$, and the random distances separating transmitters and receivers, $\{D_i\}$, on the asymptotic CCDF bounds in (\ref{eqn:s_lower})-(\ref{eqn:s}) is confined to the fractional moments $\mathbb{E}[\Psi^{\delta}]\mathbb{E}[\Psi^{-\delta}] \mathbb{E}[D^2]$.  Since the asymptotic lower bound is tight in most scenarios of interest, as explained in further detail below, the fractional moments are generally able to completely capture the effect of fading and random distances.  When channel inversion is employed, then the fractional moment dependence actually holds for the upper and lower bounds themselves, as shown in Section \ref{ssec:4b}.

{\bf Looseness of the upper bound.}  The asymptotic looseness of the upper bound depends only on the path loss exponent $\alpha$ and not on the random channel effects, i.e.,
\begin{equation}
\label{eqn:u}
\lim_{y \to \infty} \frac{\bar{F}_Y^u(y)}{\bar{F}_Y(y)} = \lim_{y \to \infty} \frac{\bar{F}_Y^u(y)}{\bar{F}_Y^l(y)} = \frac{2}{2-\delta} = \frac{\alpha}{\alpha-1}.
\end{equation}
Moreover, the upper bound is increasingly tight as $\alpha$ increases.  The fact that the upper bound, which is based on the Chebychev inequality, is not tight suggests the use of tighter upper bounds such as the Chernoff bound.  This is in fact a viable approach in theory, although it is often not computationally feasible.  An upper bound using the Chernoff bound instead of the Chebychev bound is developed in the Appendix, along with a discussion of the associated computational obstacles.

{\bf Tightness of the lower bound.}  The lower bound is tight as $y \to \infty$, i.e., as one moves further along the tail of the distribution of $Y$ (also corresponding to $\beta \to 0$).  The lower bound captures the probability of outage being caused by one or more {\em individually} dominant interferers, and thus ignores the probability that there is no single dominant interferer but
the {\em aggregate} interference level summed over all interferers causes an outage.  As a result, the fact that the lower bound is tight as $y \to \infty$ is intuitive given the fact that the distribution  $F_W$ of the channel $W = \Psi D^{-\alpha}$ is a subexponential distribution, a sub-class of heavy tailed distributions \cite{GolKlu1997}.  A key property of a subexponential distribution is that with high probability sums of subexponential random variables achieve large values by individual terms in the sum being large:
\begin{equation}
\label{eqn:v}
\lim_{x \to \infty} \frac{\mathbb{P}(X_1 + \cdots + X_n > x)}{\mathbb{P}(\max\{X_1,\ldots,X_n\} > x)} = 1, ~ n \geq 2.
\end{equation}
In the present context, as $\beta$ decreases (or equivalently, as $y$ increases) it is increasingly unlikely that a group of interferers could collaboratively cause an outage for the reference receiver without at least one of them being a dominant interferer. In most scenarios of interest, the desired outage probability is quite low and therefore $y$ is sufficiently large.  As a result, the asymptotic lower bound in (\ref{eqn:s_lower}) is generally very accurate.  The SIR threshold can also be reduced through spreading (e.g., direct sequence code division multiple access) or coding.  The impact of spreading on the outage probability (and transmission capacity) is addressed in \cite{WebYan2005} where the SIR requirement is reduced by the spreading factor.

We now utilize the results of Theorem \ref{thm:2} to generate bounds on the performance metrics of interest.  Under randomized transmission, each potential transmitter $i \in \Pi$ transmits at random (with fixed power) with a specified probability $p$.  In this case the intensity of attempted transmissions (the intensity of $\Phi$) is $\mu = \lambda p$.

\begin{theorem}
\label{thm:3}
{\em Under randomized transmissions and without channel inversion, the bounds on the outage probability (\ref{eqn:i}) are:
\begin{eqnarray}
q^u(\mu) &=& 1 - \mathbb{E}\left[\left(1 - \frac{\frac{\delta}{2-\delta} \Theta \mu} {(1 - \frac{\delta}{1-\delta} \Theta \mu)^2}
\right)^+ e^{-\Theta \mu}\right], \nonumber \\
q^l(\mu) &=& 1 - \mathbb{E}[e^{-\Theta \mu}], \label{eqn:x}
\end{eqnarray}
where
\begin{equation}
\label{eqn:w}
\Theta = K \beta^{\delta}= \pi \mathbb{E}[\Psi^{\delta}] \Psi^{-\delta} D^2 \beta^{\delta},
\end{equation}
and $\Psi \sim F_{\Psi}$ and $D \sim F_D$.  The bounds on the spatial throughput (\ref{eqn:k}) are:
\begin{eqnarray}
\tau^l(\mu) &=& \mu \mathbb{E}\left[\left(1 - \frac{\frac{\delta}{2-\delta} \Theta \mu} {(1 - \frac{\delta}{1-\delta} \Theta \mu)^2}
\right)^+ e^{-\Theta \mu}\right], \nonumber \\
\tau^u(\mu) &=& \mu \mathbb{E}[e^{-\Theta \mu}]. \label{eqn:y}
\end{eqnarray}
The bounds on the transmission capacity (\ref{eqn:l}) are
\begin{eqnarray}
c^l(\epsilon) &=& q^{u,-1}(\epsilon)(1-\epsilon), \nonumber \\
c^u(\epsilon) &=& q^{l,-1}(\epsilon)(1-\epsilon), \label{eqn:z}
\end{eqnarray}
where $q^{u,-1},q^{l,-1}$ are the inverses of $q^u,q^l$ in (\ref{eqn:x}).}
\end{theorem}

The expressions in the theorem are easily obtained by substituting the bounds on $\bar{F}_Y(y)$ from Theorem \ref{thm:2} into the performance metric expressions for $q(\mu)$ (\ref{eqn:i}), $\tau(\mu)$ (\ref{eqn:k}), and $c(\epsilon)$ (\ref{eqn:l}).   A discussion of Theorem \ref{thm:3} is found after Corollary \ref{thm:3} in Section \ref{ssec:4b}, which gives the analogous results when channel inversion is employed.

%%%%%%%%%%%%%%%%%%%%%%%%%%%%%%%%%%%%%%%%%%%%%%%%%%%%%%%%%%%%%%%%%%%%%%%
% 4-B. Performance without threshold scheduling and with channel inversion
%%%%%%%%%%%%%%%%%%%%%%%%%%%%%%%%%%%%%%%%%%%%%%%%%%%%%%%%%%%%%%%%%%%%%%%
\subsection{\label{ssec:4b}Performance without threshold scheduling and with channel inversion}

In this paper we consider two distinct ways in which CSI may be exploited by the transmitter: threshold scheduling of transmissions and channel inversion.  Channel inversion is a specific type of power control in which the transmitted power is an inverse function of the channel quality.  This is by far the most prevalent form of power control in current wireless networks.  Although fast channel inversion is a widely known feature of CDMA cellular networks for avoiding the near-far problem, channel inversion is also used in all cellular networks (sometimes called Automatic Gain Control) and also in the Bluetooth ad hoc networking standard to adjust for transmission range and channel quality.   Therefore, in this section we consider performance without threshold scheduling but with channel inversion. Performance with threshold scheduling but without channel inversion is discussed in Section \ref{ssec:5a}, and performance with both threshold scheduling and channel inversion is discussed in Section \ref{ssec:5b}.

Each transmitter $i \in \Phi$ that elects to transmit employs transmit power $P_i = \frac{1}{W_i}$, where $W_i = \Psi_{ii} D_i^{-\alpha}$ is the channel gain separating transmitter and receiver $i$; this ensures the signal power at receiver $i$ is unity.\footnote{A sufficient condition for channel inversion to require finite power almost surely is that the support of $W$ exclude the interval $[0,\epsilon)$ for some $\epsilon > 0$.  A necessary condition for finite {\em average} power is that $\mathbb{E}[1/W] < \infty$.  For some distributions, such as Rayleigh fading, the quantity $\mathbb{E}[1/W]$ is actually infinite.  The analytical results still hold in this scenario, but this condition clearly makes channel inversion impractical. However, in Section \ref{ssec:5b} we combine channel inversion with a minimum fading threshold, so that channel inversion is feasible for essentially any distribution.}

Under channel inversion, the normalized aggregate interference seen at the reference receiver is
\begin{equation}
\label{eqn:aa}
Y = \sum_{i \in \Phi} \frac{1}{W_i} \Psi_{i0} |X_i|^{-\alpha}, ~ W_i = \Psi_{ii} D_i^{-\alpha}.
\end{equation}
Because transmission decisions are made randomly, it follows that the $W_i$'s are iid according to distribution $F_W$, which is also the distribution of $W_0$ in the case of no channel inversion.

In the case of no channel inversion, the normalized interference contribution of every interferer is divided by $W_0$, the coefficient describing the channel fade and the distance-based path loss between the reference TX and RX.  As a result, the reference RX is very sensitive to the value of $W_0$.  When channel inversion is used, the normalized contribution of each interferer is divided by a {\em different} $W_i$, namely its own effective channel coefficient.  Therefore, channel inversion completely eliminates sensitivity to $W_0$, which does not even appear in (\ref{eqn:aa}), but instead introduces sensitivity to the effective channel coefficients $W_i$ of the interfering nodes.

The analysis with channel inversion is very similar to that without channel inversion, and the following corollaries are the analogues of Theorems \ref{thm:1}, \ref{thm:2}, and \ref{thm:3} for randomized transmissions with channel inversion.

\begin{corollary}[to Theorem \ref{thm:1}]
\label{cor:1} {\em  Under randomized transmissions with channel inversion, the random variable $Y$ in (\ref{eqn:aa}) is symmetric stable with characteristic function given by (\ref{eqn:b}), with stability parameter $\delta = 2/\alpha < 1$ and dispersion parameter $\gamma$ given by (\ref{eqn:o}) with $\kappa(w)$ replaced with $\kappa$ in (\ref{eqn:t}).}
\end{corollary}

The corollary follows from Theorems 1, 2, and 3 in Ilow and Hatzinakos \cite{IloHat1998}.

\begin{corollary}[to Theorem \ref{thm:2}]
\label{cor:2}
{\em Under randomized transmissions with channel inversion, the expressions $\bar{F}_Y^u,\bar{F}_Y^l$ are upper and lower bounds on the CCDF $\bar{F}_Y(y)$ of the random variable $Y$ in (\ref{eqn:aa}):
\begin{eqnarray}
\bar{F}_Y^u(y) &=& 1 - \left(1 - \frac{\frac{\delta}{2-\delta} \kappa \mu y^{-\delta}} {(1 - \frac{\delta}{1-\delta} \kappa \mu y^{-\delta})^2}
\right)^+ e^{-\kappa \mu y^{-\delta}}, \nonumber \\
\bar{F}_Y^l(y) &=& 1-e^{-\kappa \mu y^{-\delta}}, \label{eqn:ab}
\end{eqnarray}
where $\kappa$ is given in (\ref{eqn:t}).  The upper bound is nontrivial for all $\kappa \mu y^{-\delta} < h(\delta)$, defined as
\begin{equation}
\label{eqn:ac}
h(\delta) = \frac{1}{2(2-\delta)}\left[ \frac{(1-\delta)(5-3\delta)}{\delta} - \sqrt{\delta (9-5\delta)} \right].
\end{equation}
The lower bound is asymptotically tight as $y \to \infty$ and the upper bound has an asymptotic bounded error.  Specifically, $\bar{F}_Y^l(y),\bar{F}_Y(y),\bar{F}^u(y)$ have asymptotic expansions given in (\ref{eqn:s_lower})-(\ref{eqn:s}).}
\end{corollary}

The proof is found in the Appendix. The bounds on the CCDF of $F_Y$ in Corollary \ref{cor:2} may be used to obtain performance bounds for $q(\mu)$ (\ref{eqn:i}), $\tau(\mu)$ (\ref{eqn:k}), and $c(\epsilon)$ (\ref{eqn:l}), as shown in the following corollary.

\begin{corollary}[to Theorem \ref{thm:3}]
\label{cor:3}
{\em Under randomized transmissions and with channel inversion, the bounds on the outage probability (\ref{eqn:i}) are:
\begin{eqnarray}
q^u(\mu) &=& 1 - \left(1 - \frac{\frac{\delta}{2-\delta} \theta \mu} {(1 - \frac{\delta}{1-\delta} \theta \mu)^2}
\right)^+ e^{-\theta \mu}, \nonumber \\
q^l(\mu) &=& 1 - e^{-\theta \mu}, \label{eqn:ad}
\end{eqnarray}
where
\begin{equation}
\label{eqn:ae}
\theta = \mathbb{E}[\Theta] = \kappa \beta^{\delta}= \pi \mathbb{E}[\Psi^{\delta}] \mathbb{E}[\Psi^{-\delta}] \mathbb{E}[D^2] \beta^{\delta},
\end{equation}
and $\Psi \sim F_{\Psi}$ and $D \sim F_D$.  The bounds on the spatial throughput (\ref{eqn:k}) are:
\begin{eqnarray}
\tau^l(\mu) &=& \mu \left(1 - \frac{\frac{\delta}{2-\delta} \theta \mu} {(1 - \frac{\delta}{1-\delta} \theta \mu)^2}
\right) e^{-\theta \mu}, \nonumber \\
\tau^u(\mu) &=& \mu e^{-\theta \mu}. \label{eqn:af}
\end{eqnarray}
The bounds on the transmission capacity (\ref{eqn:l}) are
\begin{eqnarray}
c^l(\epsilon) &=& q^{u,-1}(\epsilon)(1-\epsilon), \nonumber \\
c^u(\epsilon) &=& \frac{-(1-\epsilon)\log(1-\epsilon)}{\theta}, \label{eqn:ag}
\end{eqnarray}
where $q^{u,-1}$ is the inverse of $q^u$ in (\ref{eqn:ad}).}
\end{corollary}
Comparing Corollaries \ref{cor:1}, \ref{cor:2}, and \ref{cor:3} with their corresponding Theorems \ref{thm:1}, \ref{thm:2}, and \ref{thm:3}, it is apparent that the primary impact of channel inversion is to remove the need to condition on the received signal power (which is unity under channel inversion).  Comparing Theorem \ref{thm:1} and Corollary \ref{cor:1}, adding channel inversion means the {\em unconditioned} distribution $F_Y$ is stable (instead of the conditioned distribution $F_{Y|W}$), and the dispersion parameter is given by constant $\kappa$ in (\ref{eqn:t}) instead of the function $\kappa(w)$ in (\ref{eqn:p}).  Note that $\kappa(w) = \mathbb{E}[K] = \mathbb{E}[\kappa(W)]$.  In Theorem \ref{thm:2} the bounds on the CCDF $\bar{F}_Y$ are expressed in terms of expectations of functions of  the random variable $K$ in (\ref{eqn:r}); in Corollary \ref{cor:2} the bounds on the CCDF $\bar{F}_Y$ are expressed in terms of the same functions, with $K$ replaced by its expected value, $\kappa = \mathbb{E}[K]$.  A similar comment holds for Theorem \ref{thm:3} and Corollary \ref{cor:3}. Note that the bounds in Theorems \ref{thm:2} and \ref{thm:3} require evaluating an integral, while the bounds in Corollaries \ref{cor:2} and \ref{cor:3} only require evaluating a constant.

The intuition for this difference is quite straightforward.  Without channel inversion, the marks of the Poisson process in (\ref{eqn:m}) are $\frac{\Psi_{i0}}{W_0}$ and are not independent because $W_0$ appears in each term.  As a result, the distribution of $Y$ conditioned on $W_0$ must be considered, which results in an additional expectation in the associated bounds.  With power
control, the marks of the Poisson process in (\ref{eqn:aa}) are $\frac{\Psi_{i0}}{W_i}$ and thus are independent.

%%%%%%%%%%%%%%%%%%%%%%%%%%%%%%%%%%%%%%%%%%%%%%%%%%%%%%%%%%%%%%%%%%
\subsection{\label{ssec:4d}Discussion}

In this section we discuss the preceding analytical results by comparing performance with and without channel inversion as well as studying the effect of channel fading and random distances on ad hoc network performance.

\noindent{\bf The effect of channel inversion.}  By applying Jensen's inequality to the convex function $e^{-\theta \mu}$, we can order the outage probability lower bounds in Theorem and Corollary \ref{thm:3} as:
\begin{equation}
q^l_{\rm npc}(\mu) = 1 - \mathbb{E}[e^{-\Theta \mu}] < 1 - e^{-\mathbb{E}[\Theta]\mu} = q^l_{\rm pc}(\mu),
\end{equation}
where npc and pc denote no power control and power control respectively.  Thus channel inversion strictly increases the lower bound on outage probability.  The intuition for this increase appears to come from the difference in the normalized interference expressions with and without channel inversion in (\ref{eqn:aa}) and (\ref{eqn:m}), respectively. With channel inversion, the reference receiver is vulnerable to signal fades of {\em any} of its nearby interferers (i.e., small values of $W_i$); without channel inversion, the reference receiver is vulnerable only to a fade on its own channel $W_0$.  Channel inversion introduces an undesirable diversity on the interference power that increases the likelihood of a nearby dominant interferer causing an outage. Numerical results indicate that similar conclusions hold for the actual outage probability, not just for the analytical bounds.

There are a few other relevant issues concerning channel inversion that should also be mentioned.  If channel inversion is used the average transmission power is $\mathbb{E}[1/W]$.  An equivalent fixed power system that delivers the same average received power would only require transmission power of $\frac{1}{\mathbb{E}[W]}$, which by Jensen's inequality is smaller than $\mathbb{E}[1/W]$.  Thus, channel inversion essentially requires greater transmission power, or alternatively delivers less received power, than a system using fixed power.  As a result, using channel inversion has the potential of pushing it from the interference-limited regime into the noise-limited regime.  This effect does not appear in our SIR-based analysis, but we note that this effect is less pronounced when channel inversion is combined with threshold scheduling in Section \ref{ssec:5b} because the threshold eliminates small values of $W_i$ and thus decreases the difference between $\mathbb{E}[1/W]$ and $\frac{1}{\mathbb{E}[W]}$.

One positive effect of channel inversion is that it assists with fairness.  If the distance between a transmitter-receiver pair is large compared to the average and/or the channel gain coefficient is small, the outage probability of this pair would be considerably higher than the network-wide average without channel inversion.  Channel inversion neutralizes distance and/or fading disadvantages, and essentially puts all transmitter-receiver pairs on equal footing.  What our results show is that there is a quantifiable network-wide penalty for doing so.\\

\noindent{\bf Effect of random distance and fading.} In order to understand the effect of random Tx-Rx distance and fading, it is useful to rewrite the expression for the transmission capacity upper bound $c^u(\epsilon)$ for a power-controlled system given in
(\ref{eqn:ag}):
\begin{eqnarray}
c^u(\epsilon) &=& \frac{-(1-\epsilon)\log(1-\epsilon)}{\theta}
\approx \frac{ \epsilon (1- \epsilon)}{\theta} \nonumber \\
& = & \frac{\epsilon (1- \epsilon) \beta^{-\delta}}
{\pi \mathbb{E}[\Psi^{\delta}] \mathbb{E}[\Psi^{-\delta}] \mathbb{E}[D^2]},\label{eq:tc-approx}
\end{eqnarray}
where we have used $-\log(1 - \epsilon) \approx \epsilon$ for small values of $\epsilon$.\footnote{When there is no fading ($\Psi_{ij} = 1$), transmitter to receiver distances are fixed ($D_i = r$), and $\epsilon$ is small, (\ref{eq:tc-approx}) recovers the TC given in Theorem 1 in \cite{WebYan2005}, which describes the TC of a network in which there is only path-loss.}  Although this bound holds for channel inversion power controlled systems, it is also extremely accurate for systems using fixed transmission power when $\epsilon$ is small because the asymptotically tight (i.e., for $\epsilon \to 0$) outage lower bound given in (\ref{eqn:s_lower}) leads to the same transmission capacity upper bound stated above.

\underline{Channel variations reduce transmission capacity.}
Applying Jensen's inequality to the convex function $\frac{1}{x}$ and random variable $\Psi^{\delta}$ yields
\begin{equation}
\label{eqn:aj}
\mathbb{E}[\Psi^{\delta}] \mathbb{E}[\Psi^{-\delta}] \geq 1,
\end{equation}
with equality iff $\Psi$ is deterministic. As a result, $\theta$ in (\ref{eqn:ae}) is strictly larger under random $\Psi$ than under deterministic $\Psi$, and thus fading reduces transmission capacity under randomized transmissions with channel inversion.  It can also be seen that variations in the distances separating transmitters and receivers also reduces transmission capacity.  The ratio of $\theta$ with variable $\{D_i\}$ over $\theta$ with fixed $\{D_i\}$ (with the same mean) is given by
\begin{equation}
\label{eqn:ak}
\frac{\mathbb{E}[D^2]}{\mathbb{E}[D]^2} = 1 + \frac{{\rm Var}[D]}{\mathbb{E}[D]^2},
\end{equation}
the right hand term is the factor by which $\theta$ increases due to random distances, and thereby reduces the transmission capacity.

\underline{Separating the effects of signal and interference fading.} The effect of fading can be more clearly elucidated by separating signal and interference fading:

{\em Corollary: Under randomized transmissions with channel inversion, if the reference channel gain $\Psi_{00}$ is drawn according to distribution $F_{\Psi_S}$ ($S$ for signal) while the interference channel gains $\{\Psi_{i0}\}$ are drawn (iid) according to a possibly different distribution $F_{\Psi_I}$ ($I$ for interference), all results of Corollaries \ref{cor:1}, \ref{cor:2}, and \ref{cor:3} hold with $\kappa$ in (\ref{eqn:t}) defined as:
\begin{equation}
\label{eqn:al}
\kappa = \pi \mathbb{E}[\Psi_I^{\delta}] \mathbb{E}[\Psi_S^{-\delta}] \mathbb{E}[D^2].
\end{equation}}

This statement follows from the fact that the proof of Corollary \ref{cor:2} only depends on the $\delta$-moment of $Z =
\frac{\Psi_{i0}}{\Psi_{00}} D_0^{\alpha}$ given by $\mathbb{E}[Z^{\delta}] = \mathbb{E}[\Psi_{i0}^{\delta}] \mathbb{E}[\Psi_{00}^{-\delta}] \mathbb{E}[D_0^2]$, and thus does not require $\Psi_{00}$ and $\Psi_{i0}$ to follow the same distribution.

To simplify discussion, without loss of generality assume $\mathbb{E}[\Psi_S] = \mathbb{E}[\Psi_I] = 1$.  Since the functions $x^{-\delta}$ and $x^{\delta}$ are convex and concave, respectively, for $0 < \delta < 1$, Jensen's inequality yields $\mathbb{E}[\Psi_S^{-\delta}] \geq \mathbb{E}[\Psi_S]^{-\delta} = 1$ and $\mathbb{E}[\Psi_I^{\delta}] \leq \mathbb{E}[\Psi_I]^{\delta} = 1$. As a
result, fading of the desired signal {\em reduces} transmission capacity, while fading of interfering signals {\em increases} transmission capacity.

Since the function $x^{-\delta}$ approaches infinity as $x \rightarrow 0$, if $F_{\Psi_S}$ has a large amount of mass near the origin the fractional moment $\mathbb{E}[\Psi_S^{-\delta}]$ can be very large, thereby leading to a significant reduction in transmission capacity.  The proceeding section shows that this indeed the case when $\Psi_S$ is exponential (i.e., Rayleigh fading), where $\mathbb{E}[\Psi_S^{-\delta}] \rightarrow \infty$ as $\delta \rightarrow 1$.

There are distributions for which the fractional moment $\mathbb{E}[\Psi_I^{\delta}]$ can be made arbitrarily small, implying an arbitrarily large increase in transmission capacity, but this is certainly the exception.  A simple calculation shows that $\mathbb{E}[\Psi_I^{\delta}]$ is lower bounded by the probability $\Psi_I$ is greater than or equal to unity (i.e., the mean), which is reasonably large for typical fading distributions.

Therefore, it is safe to say that signal fading can have a rather significant negative effect on transmission capacity, while interference fading leads to a less significant positive effect.  If the fading distributions are identical, then (\ref{eqn:aj}) indicates that the net effect of fading is negative.

\underline{Maximum achievable spatial throughput and TC.} The optimal transmission probability to maximize the spatial throughput upper bound $\tau^u(\lambda p)$ in (\ref{eqn:af}) is $p^{u,*} = \frac{1}{\lambda \theta}$, and thus the bound optimal intensity of transmission attempts is $\mu^{u,*} = \frac{1}{\theta} \land \lambda$.  The corresponding bound on optimal throughput is
\begin{equation}
\label{eqn:ah}
\tau^{u,*} = \tau^u(\mu^{u,*}) = \frac{1}{e \theta}, ~ \lambda > 1/\theta.
\end{equation}
The outage probability constraint that maximizes the transmission capacity upper bound $c^u(\epsilon)$ in (\ref{eqn:ag}) is
$\epsilon^{u,*} = 1 - 1/e \approx 0.63$, with a corresponding bound on transmission capacity of
\begin{equation}
\label{eqn:ai}
c^{u,*} = c^u(\epsilon^{u,*}) = \frac{1}{e \theta} = \tau^{u,*}, ~ \lambda > 1/\theta.
\end{equation}
Note that maximizing the spatial throughput and TC incurs a potentially unacceptable high outage probability: almost two thirds of all attempted transmissions must fail to maximize capacity.  The assumption that $\lambda > 1/\theta$ means that the spatial intensity of potential transmitters in $\Pi$ is sufficiently large to ``saturate'' the network, i.e., the network will not be under-utilized due to a lack of available transmitters.  We emphasize that the optimality of $p^{u,*},\mu^{u,*}, \epsilon^{u,*}$ holds for the bounds, not the performance metric itself, however our numerical and simulation results will shown that the approximation is valid over most regimes of interest.

%%%%%%%%%%%%%%%%%%%%%%%%%%%%%%%%%%%%%%%%%%%%%%%%%%%%%%%%%%%%%%%%%%%%%%%
% 4-C. Examples
%%%%%%%%%%%%%%%%%%%%%%%%%%%%%%%%%%%%%%%%%%%%%%%%%%%%%%%%%%%%%%%%%%%%%%%
\subsection{\label{ssec:4c}Examples}

We next compute the performance bounds in Theorem \ref{thm:3} and Corollary \ref{cor:3} for three examples.  In Example \ref{exa:1} we fix the transmitter to receiver distances, $D_i = r$, and let $\Psi$ have a lognormal distribution, capturing the impact of lognormal shadowing.  In Example \ref{exa:2} we again fix $D_i = r$ and let $\Psi$ have an exponential distribution, capturing the
impact of Rayleigh fading.  In Example \ref{exa:3} we fix the fading coefficients $\Psi_{ij} = 1$ and let $D$ have a distribution corresponding to the distance to the nearest neighbor in a Poisson process of potential receivers.

\begin{example}
\label{exa:1} {\bf Lognormal shadowing.}  Fix $D_i = r$ for each $i$, and let $\Psi$ be lognormal distributed with parameter $\sigma^2$, i.e., $\Psi \sim \mathcal{LN}(0,\sigma^2)$, where
\begin{equation}
\label{eqn:am}
f_{\Psi}(\psi) = \frac{1}{\sqrt{2 \pi \sigma^2} \psi} \exp \left\{-\frac{(\log \psi)^2}{2 \sigma^2} \right\}, ~ \psi > 0.
\end{equation}
Under randomized transmissions in the absence of channel inversion the distribution of received signal power is
\begin{equation}
\label{eqn:an}
F_W(w) = F_{\Psi}(w r^{\alpha}) = 1-Q\left(\frac{\log (w r^{\alpha})}{\sigma}\right),
\end{equation}
where $Q(\cdot)$ is the CCDF of the $\mathcal{N}(0,1)$ standard normal distribution.  It is straightforward to establish that:
\begin{equation}
\label{eqn:ao}
\mathbb{E}[\Psi^{\delta}] = \mathbb{E}[\Psi^{-\delta}] = \exp \left\{ \frac{\delta^2 \sigma^2}{2} \right\},
\end{equation}
so that
\begin{equation}
\label{eqn:ap}
\kappa(w) = \pi \exp \left\{ \frac{\delta^2 \sigma^2}{2} \right\} w^{-\delta}, ~~ \kappa = \pi \exp \{ \delta^2 \sigma^2 \} r^2.
\end{equation}
The bounds on $q(\mu)$ in (\ref{eqn:x}) are
\begin{eqnarray}
q^u(\mu) &=& 1 - \int_0^{\infty} \left(1 - \frac{\frac{\delta}{2-\delta} a \mu \psi^{-\delta}} {(1 - \frac{\delta}{1-\delta} a \mu \psi^{-\delta})^2}
\right) \times \nonumber \\
& & \exp \{-a \mu \psi^{-\delta}\} {\rm d}F_{\Psi}(\psi), \nonumber \\
q^l(\mu) &=& 1 - \int_0^{\infty} \exp \{-a \mu \psi^{-\delta}\} {\rm d}F_{\Psi}(\psi), \label{eqn:aq}
\end{eqnarray}
for
\begin{equation}
\label{eqn:ar}
a = \pi \exp \left\{ \frac{\delta^2 \sigma^2}{2} \right\} r^2 \beta^{\delta}.
\end{equation}
The bounds on $\tau(\mu)$ and $c(\epsilon)$ in Theorem \ref{thm:3} are computable from the bounds on $q(\mu)$.  All quantities in Corollaries \ref{cor:1}, \ref{cor:2}, \ref{cor:3} are computable from $\kappa$ given above.
\end{example}

\begin{example}
\label{exa:2} {\bf Rayleigh fading.}  Fix $D_i = r$ for each $i$, and let $\Psi$ be exponentially distributed, i.e., $\Psi \sim {\rm Exp}(1)$.  Under randomized transmissions in the absence of channel inversion the distribution of received signal power is also exponential with parameter $r^{\alpha}$, i.e.,
\begin{equation}
\label{eqn:as}
F_W(w) = F_{\Psi}(w r^{\alpha}) = 1 - \exp \{ - w r^{\alpha} \}.
\end{equation}
It is straightforward to establish that
\begin{equation}
\label{eqn:at}
\mathbb{E}[\Psi^{\delta}] = \Gamma(1+\delta), ~~
\mathbb{E}[\Psi^{-\delta}] = \Gamma(1-\delta),
\end{equation}
so that
\begin{equation}
\label{eqn:au}
\kappa = \pi \frac{\pi \delta}{\sin(\pi \delta)} r^2,
\end{equation}
where we have used $\Gamma(1+\delta) \Gamma(1-\delta) = \frac{\pi \delta}{\sin(\pi \delta)}$.

If we plug this value of $\kappa$ into the lower bound to outage probability with channel inversion in (\ref{eqn:ad}), we get:
\begin{eqnarray}
q^l(\mu) = 1 - \exp \{- \pi \mu r^2 \Gamma(1 + \delta) \Gamma(1 - \delta) \}.
\label{eq-rayleigh_exact}
\end{eqnarray}
In \cite{BacBla2006}, a closed form expression (Corollary 3.2) for the outage probability of a system utilizing random transmissions {\em without} channel inversion in a Rayleigh fading environment is derived using the moment generating function of the interference power.  Remarkably, this expression coincides {\em exactly} with the above expression, which is an outage probability lower bound when there is channel inversion.  As a result, (\ref{eq-rayleigh_exact}) corresponds to the exact outage probability without channel inversion, and we can unequivocally state that the use of channel inversion degrades performance in Rayleigh fading, since (\ref{eq-rayleigh_exact}) is an outage lower bound when channel inversion is used.

When no channel inversion is employed, we can translate (\ref{eq-rayleigh_exact}) into exact expressions for the other performance metrics; these expressions are all upper bounds to performance with channel inversion:
\begin{eqnarray}
c(\epsilon) &=& - \frac{\beta^{-\delta} r^{-2} \log(1-\epsilon) (1- \epsilon)}
{\pi \Gamma(1 + \delta) \Gamma(1 - \delta)} , \nonumber \\
\tau(\mu) &=& \mu \exp\{-\pi \mu \beta^{\delta} r^2 \Gamma(1 + \delta) \Gamma(1 - \delta)\}.
\end{eqnarray}

\begin{figure}[ht]
\centering
\includegraphics[width=3.5in]{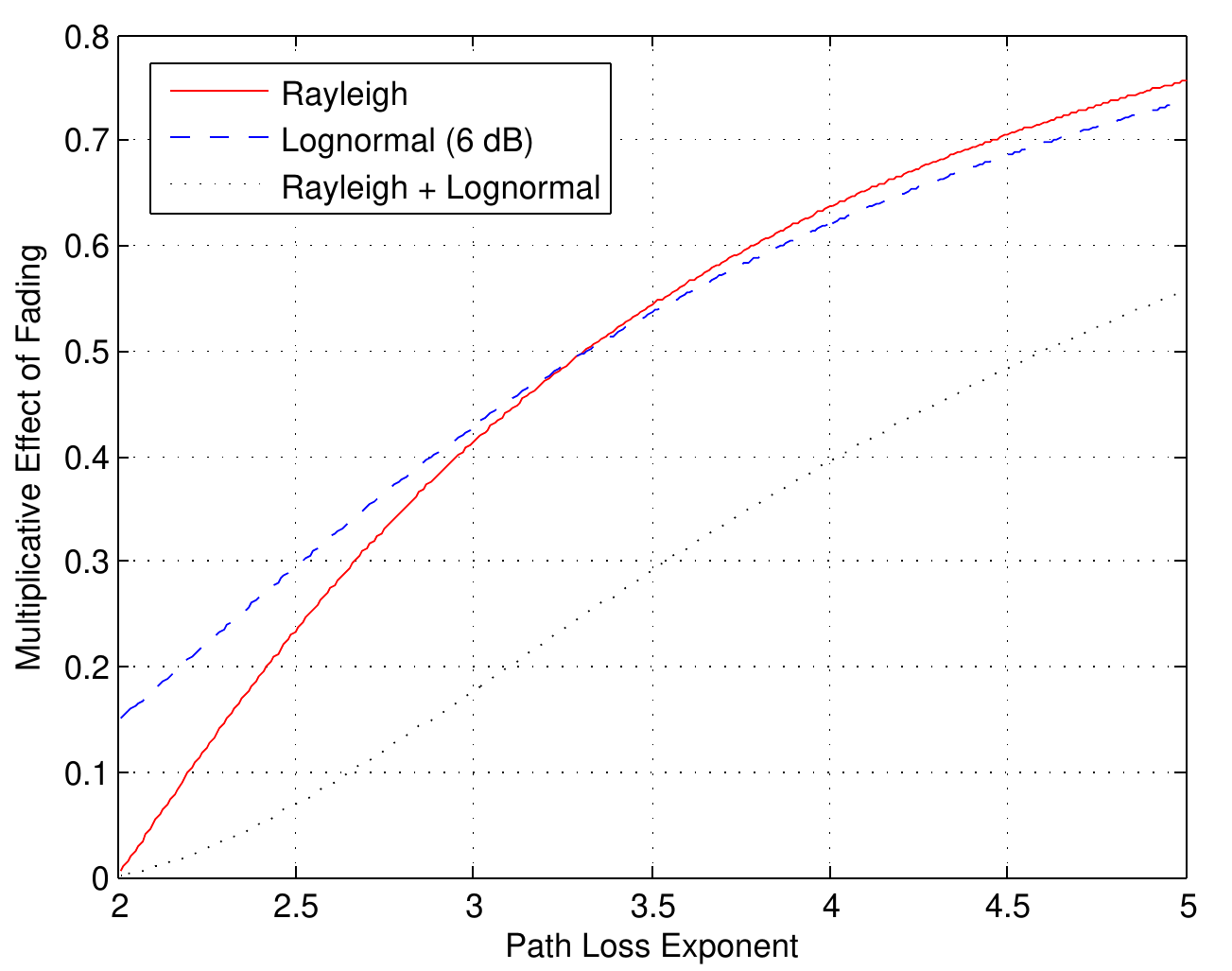}
\caption{Multiplicative effect of Rayleigh and lognormal fading.}
\label{fig-rayleigh_kappa}
\end{figure}

In Figure \ref{fig-rayleigh_kappa} the quantity $\frac{\pi r^2}{\kappa} = \frac{1}{\mathbb{E}[\Psi^{\delta}] \mathbb{E}[\Psi^{-\delta}]}$, which is the multiplicative effect of fading on transmission capacity, is plotted against the path-loss exponent $\alpha$ for Rayleigh fading, lognormal fading ($\sigma = 6$ dB), and the combination of the two.  Both fading distributions have a more benign effect as the path-loss exponent increases, but note that Rayleigh fading exacts a very harsh penalty when the path-loss exponent is near two.

\end{example}

\begin{example}
\label{exa:3} {\bf Nearest receiver transmissions.} Fix $\Psi_{ij} = \psi$ for each $i,j$. Recall $\Pi$ is the MPPP of intensity $\lambda$ of potential transmitters, where each potential transmitter $i$ has an intended receiver not in $\Pi$.  Suppose the set of all possible receivers is a PPP, denoted $\Pi'$, of intensity $\lambda'$.  Consider the case when each node elects to transmit to its nearest neighbor; for simplicity we ignore the facts that $i)$ multiple potential transmitters may select the same receiver, and $ii)$ the distances are dependent random variables.  Under these assumptions the distances, $\{D_i\}$, are iid with distribution
\begin{equation}
\label{eqn:aw}
\bar{F}_D(d) = \mathbb{P}(\Pi' \cap b(o,d) = \emptyset) = \exp\{-\pi \lambda' d^2\}.
\end{equation}
Then
\begin{equation}
\label{eqn:ax}
\mathbb{E}[D] = \frac{1}{2 \sqrt{\lambda'}}, ~~
\mathbb{E}[D^2] = \frac{1}{\pi \lambda'}.
\end{equation}
Under randomized transmissions in the absence of channel inversion the distribution of received signal power
\begin{equation}
\label{eqn:ay}
F_W(w) = \bar{F}_D((\psi/w)^{\frac{1}{\alpha}}) = \exp \{ - \pi \lambda' (\psi/w)^{\delta} \}.
\end{equation}
It follows that $\kappa(w) = \pi \psi w^{-\delta}$ and $\kappa = \frac{1}{\lambda'}$.  The bounds on $q(\mu)$ in (\ref{eqn:x}) are given by
\begin{eqnarray}
q^u(\mu) &=& 1 - \int_0^{\infty} \left(1 - \frac{\frac{\delta}{2-\delta} \pi \beta^{\delta} \mu x^2} {(1 - \frac{\delta}{1-\delta} \pi \beta^{\delta} \mu x^2)^2}
\right) \times \nonumber \\
& & \exp \{-\pi \beta^{\delta} \mu x^2\} {\rm d}F_D(x), \nonumber \\
q^l(\mu) &=& 1 - \int_0^{\infty} \exp \{-\pi \beta^{\delta} \mu x^2\} {\rm d}F_D(x) \nonumber \\
&=& \frac{\beta^{\delta} \mu}{\beta^{\delta} \mu + \lambda'}. \label{eqn:ba}
\end{eqnarray}
The lower bounds $\tau^l(\mu)$ and $c^l(\epsilon)$ in Theorem \ref{thm:3} are computable from the upper bound on $q^u(\mu)$, with the upper bounds being
\begin{equation}
\label{eqn:bb}
\tau^u(\mu) = \mu \left(1 - \frac{\beta^{\delta} \mu}{\beta^{\delta} \mu + \lambda'} \right), ~~
c^u(\epsilon) = \frac{\lambda'}{\beta^{\delta}} \epsilon.
\end{equation}
All quantities in Corollaries \ref{cor:1}, \ref{cor:2}, \ref{cor:3} are computable from $\kappa$ given above.
\end{example}

%%%%%%%%%%%%%%%%%%%%%%%%%%%%%%%%%%%%%%%%%%%%%%%%%%%%%%%%%%%%%%%%%%%%%%%
% 5. Performance with threshold scheduling
%%%%%%%%%%%%%%%%%%%%%%%%%%%%%%%%%%%%%%%%%%%%%%%%%%%%%%%%%%%%%%%%%%%%%%%
\section{\label{sec:5}Performance with threshold scheduling}

In this section we study the performance when each potential transmitter $i \in \Pi$ elects to transmit only if the channel strength to its intended receiver is acceptably strong, i.e., if $W_i = \Psi_{ii} D_i^{-\alpha} > t$, where $t$ is the global channel state threshold.  In particular, the set of actual transmitters, $\Phi$, is given by $\Phi = \{i \in \Pi : W_i > t\}$.  By the
assumed independence of signal strengths across potential transmitters, the intensity of attempted transmissions is
\begin{equation}
\label{eqn:bc}
\mu(t) = \lambda \mathbb{P}(\Psi D^{-\alpha} > t) = \lambda \bar{F}_W(t).
\end{equation}
The motivation behind this scheduling policy is the intuition that transmitting only when the channel to one's intended receiver is strong may significantly improve performance above randomized transmission decisions.  We emphasize there is no claim that the threshold scheduling rule is in any sense globally optimal: global optimality would require global channel state knowledge by each
node, which is clearly unrealistic.  Transmitter channel state information is a realistic assumption when channel coherence times extend across multiple transmission attempts, which is the case for all but the highest mobility systems. We consider performance both with fixed (unit) power (Section \ref{ssec:5a}) and with channel inversion (Section \ref{ssec:5b}).

Note that a threshold based policy is most feasible when the timescale of fading is smaller than the allowable packet delays.  If this is the case, delay constraints are not violated even if a transmitter has to wait multiple coherence times before the threshold is exceeded.  In slow fading scenarios, it may not be possible to employ only a threshold-based schemes, and some combination of randomized scheduling and threshold scheduling may be more appropriate.

%%%%%%%%%%%%%%%%%%%%%%%%%%%%%%%%%%%%%%%%%%%%%%%%%%%%%%%%%%%%%%%%%%%%%%%
% 5-A. Performance with threshold scheduling and without channel inversion
%%%%%%%%%%%%%%%%%%%%%%%%%%%%%%%%%%%%%%%%%%%%%%%%%%%%%%%%%%%%%%%%%%%%%%%
\subsection{\label{ssec:5a}Performance with threshold scheduling and without channel inversion}

In the absence of channel inversion, the normalized aggregate interference seen by the reference receiver is
\begin{equation}
\label{eqn:bd}
Y = \frac{1}{W_{0|t}} \sum_{i \in \Phi} \Psi_{i0} |X_i|^{-\alpha}, ~ W_{0|t} = \Psi_{00} D_0^{-\alpha},
\end{equation}
where $W_{0|t}$ is the received signal power conditioned on the reference transmitter having an acceptably strong channel.  Notice that the effect of the threshold policy is to change the distribution of the channel coefficient from the unconditional distribution of $W$ to the conditional distribution of $W$ given $W \geq t$; the distribution of the interfering channel coefficients $\Psi_{i0}$ is  unaffected because transmission is decided only on the basis of $\Psi_{ii}$.

The distribution of $W_{0|t}$ is then
\begin{equation}
\label{eqn:be}
W_{0|t} \sim F_{W|t} = \mathbb{P}(W \leq w ~ | ~ W > t), ~~ w \geq t,
\end{equation}
which is expressible in terms of the known distributions $F_{\Psi}$ and $F_D$.  The distribution of $Y$ may be expressed in terms of the distribution of $Y$ conditioned on $W_{0|t}$:
\begin{equation}
\label{eqn:bf}
\bar{F}_Y(y) = \int_t^{\infty} \bar{F}_{Y|W_t}(y|w_t) {\rm d}F_{W_t}(w_t),
\end{equation}
where $W_t \sim F_{W|t}$.

\begin{theorem}
\label{thm:4}
{\em Under threshold based transmissions and lacking channel inversion, Theorems \ref{thm:1}, \ref{thm:2}, and \ref{thm:3} continue to hold with the following changes.  In Theorem \ref{thm:1}, the conditional distribution $\bar{F}_{Y|W}(y|w)$ in (\ref{eqn:n}) is replaced with $\bar{F}_{Y|W_t}(y|w_t)$ in (\ref{eqn:bf}).  In Theorem \ref{thm:2} the upper and lower bounds on the CCDF $\bar{F}_Y(y)$ of the random variable $Y$ in (\ref{eqn:m}) are replaced with upper and lower bounds on the CCDF $\bar{F}_Y(y)$ of the random variable $Y$ in (\ref{eqn:bd}), with $K$ in (\ref{eqn:r}) replaced with
\begin{equation}
\label{eqn:bg}
K_t = \pi \mathbb{E}[\Psi^{\delta}] W_t^{-\delta},
\end{equation}
and $\kappa$ in (\ref{eqn:t}) replaced with
\begin{eqnarray}
\kappa(t) &=& \mathbb{E}[K_t] = \pi \mathbb{E}[\Psi^{\delta}] \mathbb{E}[W_t^{-\delta}] = \pi \mathbb{E}[\Psi^{\delta}] \frac{\mathbb{E}[W^{-\delta}\mathbf{1}_{W > t}]}{\bar{F}_W(t)} \nonumber \\
&=& \pi \mathbb{E}[\Psi^{\delta}] \frac{\int_0^{\infty} x^2 \left[ \int_{t x^{\alpha}}^{\infty} v^{-\delta} {\rm d}F_{\Psi}(v) \right] {\rm d}F_D(x)}{\int_0^{\infty} \left[ \int_{t x^{\alpha}}^{\infty} {\rm d}F_{\Psi}(v) \right] {\rm d}F_D(x)}. \label{eqn:bh}
\end{eqnarray}
In Theorem \ref{thm:3} the upper and lower bounds on the outage probability and the spatial throughput hold with $\Theta$ in (\ref{eqn:w}) replaced with
\begin{equation}
\label{eqn:bi}
\Theta_t = K_t \beta^{\delta} = \pi \mathbb{E}[\Psi^{\delta}]W_t^{-\delta} \beta^{\delta}.
\end{equation}
The bounds on the transmission capacity are replaced with the inverses of $q^u,q^l$ defined in terms of the expectation of $\Theta_t$.  The proof is found in the Appendix.}
\end{theorem}

\noindent {\bf Comments on Theorem \ref{thm:4}.}  Several sanity checks are available to validate the above expression for $\kappa(t)$.  First, note that for $t = 0$ (no signal strength threshold for transmission) $\kappa(t)$ reduces to $\kappa$.  Second, consider the case when the channel fading gains are constant, $\Psi_{ij} = \psi$ for all $i,j$.  Then it is straighforward to see that
\begin{equation}
\label{eqn:bj}
\kappa(t) = \pi \frac{\mathbb{E}[D^2 \mathbf{1}_{D < (\psi/t)^{\frac{1}{\alpha}}}]}{F_D((\psi/t)^{\frac{1}{\alpha}})}.
\end{equation}
Note that the condition $D < (\psi/t)^{\frac{1}{\alpha}}$ is equivalent to $\psi D^{-\alpha} > t$, which is the signal strength transmission requirement under fixed fading and random distances.  Finally, consider the case when the transmitter to receiver distances are constant, $D_i = r$ for all $i$.  Then
\begin{equation}
\label{eqn:bk}
\kappa(t) = \pi r^2 \mathbb{E}[\Psi^{\delta}] \frac{\mathbb{E}[\Psi^{-\delta}\mathbf{1}_{\Psi > t r^{\alpha}}]}{\bar{F}_{\Psi}(t r^{\alpha})}.
\end{equation}
Note that the condition $\Psi > t r^{\alpha}$ is equivalent to $\Psi r^{-\alpha} > t$, which is the signal strength transmission requirement under random fading and fixed distances.

%%%%%%%%%%%%%%%%%%%%%%%%%%%%%%%%%%%%%%%%%%%%%%%%%%%%%%%%%%%%%%%%%%%%%%%
% 5-B. Performance with threshold scheduling and with channel inversion
%%%%%%%%%%%%%%%%%%%%%%%%%%%%%%%%%%%%%%%%%%%%%%%%%%%%%%%%%%%%%%%%%%%%%%%
\subsection{\label{ssec:5b}Performance with threshold scheduling and with channel inversion}

Under threshold scheduling the transmitters in $\Phi$ are those potential transmitters in $\Pi$ with $W_i > t$.  With channel inversion, each transmitter $i$ employs transmit power $P_i = \frac{1}{W_i}$, meaning that the maximum transmit power is $P^{\rm max} = \frac{1}{t}$.  Under this channel inversion scheme the normalized aggregate interference seen at the reference receiver is
\begin{equation}
\label{eqn:bl}
Y = \sum_{i \in \Phi} \frac{1}{W_{i|t}} \Psi_{i0} |X_i|^{-\alpha},
\end{equation}
where each $W_{i|t}$ has distribution $F_{W|t}$ in (\ref{eqn:be}).

\begin{theorem}
\label{thm:5}
{\em Define
\begin{equation}
\label{eqn:bm}
\theta(t) = \kappa(t) \beta^{\delta},
\end{equation}
with $\kappa(t)$ defined in (\ref{eqn:bh}).  Under threshold based transmissions with channel inversion, Corollaries \ref{cor:1}, \ref{cor:2}, and (\ref{eqn:ad}) and (\ref{eqn:af}) in Corollary \ref{cor:3} continue to hold with $(\kappa,\theta,\mu)$ replaced with $(\kappa(t),\theta(t),\mu(t))$.  The bounds on the transmission capacity are
\begin{eqnarray}
c_l(\epsilon) &=& \lambda \bar{F}_W(\gamma^{-1}(q^{u,-1}(\epsilon))), ~
\epsilon \in [0,q^u(\theta \lambda)], \nonumber \\
c_u(\epsilon) &=& \lambda \bar{F}_W(\gamma^{-1}(q^{l,-1}(\epsilon))), ~
\epsilon \in [0,q^l(\theta \lambda)], \label{eqn:bn}
\end{eqnarray}
where $\gamma^{-1}$ is the inverse of $\gamma(t) = \theta(t) \mu(t)$, $q^{l,-1}(\epsilon) = - \log(1-\epsilon)$ and $q^{u,-1}(\epsilon)$ are the inverses of $q^l,q^u$ in (\ref{eqn:ad}) respectively, and $\theta$ is given by (\ref{eqn:ae}).  The proof is found in the Appendix.}\\
\end{theorem}

\noindent {\bf Comments on Theorem \ref{thm:5}.} By construction both $\tau(\mu)$ and $c(\epsilon)$ are concave functions, and have the same maximum value.  We can write:
\begin{eqnarray}
\tau(t) &=& \mu(t) \exp \{ - \mu(t) \theta(t) \} \nonumber \\
&=& \lambda \bar{F}_W(t) \exp \{ - \pi \lambda \beta^{\delta} \mathbb{E}[\Psi^{\delta}] \mathbb{E}[W^{-\delta}\mathbf{1}_{W > t}] \}.
\end{eqnarray}
The derivative is:
\begin{eqnarray}
\frac{{\rm d}}{{\rm d}t} \tau(t) &=& \lambda \exp \{-\pi \lambda \beta^{\delta} \mathbb{E}[\Psi^{\delta}] \mathbb{E}[W^{-\delta}\mathbf{1}_{W > t}] \} \times \nonumber \\
& & \left( \frac{{\rm d}}{{\rm d}t} \bar{F}_W(t) - a \bar{F}_W(t) \frac{{\rm d}}{{\rm d}t} \mathbb{E}[W^{-\delta}\mathbf{1}_{W > t}] \right),
\end{eqnarray}
for $a = \pi \lambda \beta^{\delta} \mathbb{E}[\Psi^{\delta}]$.
Note that
\begin{equation}
\frac{{\rm d}}{{\rm d}t} \bar{F}_W(t) = - f_W(t), ~~
\frac{{\rm d}}{{\rm d}t} \mathbb{E}[W^{-\delta}\mathbf{1}_{W > t}] = -t^{-\delta} f_W(t),
\end{equation}
so that the sufficient condition for optimality is
\begin{equation}
\frac{{\rm d}}{{\rm d}t} \tau(t) = 0 ~ \Leftrightarrow ~ \bar{F}_W(t) = \frac{t^{\delta}}{a}.
\end{equation}
The optimal throughput can be expressed as
\begin{equation}
\tau(t_{\rm opt}) = \frac{\lambda}{a} t_{\rm opt}^{\delta} \exp \{- t_{\rm opt}^{\delta} \mathbb{E}[W^{-\delta}\mathbf{1}_{W > t_{\rm opt}}] \}.
\end{equation}

The function $\kappa(t)$ is monotonically decreasing in $t$.  Taking the derivative yields:
\begin{eqnarray}
\frac{{\rm d} }{{\rm d}t} \kappa(t)
&=& \frac{\pi \mathbb{E}[\Psi^{-\delta}]}{\bar{F}_W(t)^2} \left(\bar{F}_W(t) \frac{{\rm d}}{{\rm d}t}\mathbb{E}[W^{-\delta}\mathbf{1}_{W > t}] - \right. \nonumber \\
& & \left. \mathbb{E}[W^{-\delta}\mathbf{1}_{W > t}] \frac{{\rm d}}{{\rm d}t} \bar{F}_W(t)\right) \nonumber \\
&=& \frac{\pi \mathbb{E}[\Psi^{-\delta}] f_W(t)}{\bar{F}_W(t)^2} \left( \mathbb{E}[W^{-\delta}\mathbf{1}_{W > t}] - t^{-\delta} \bar{F}_W(t) \right) \nonumber \\
&=& \frac{\pi \mathbb{E}[\Psi^{-\delta}] f_W(t)}{\bar{F}_W(t)^2} \int_t^{\infty} (w^{-\delta} - t^{-\delta}) {\rm d}F_W(t) \nonumber \\
& < & 0.
\end{eqnarray}

%%%%%%%%%%%%%%%%%%%%%%%%%%%%%%%%%%%%%%%%%%%%%%%%%%%%%%%%%%%%%%%%%%%%%%%
% 5-C. Examples
%%%%%%%%%%%%%%%%%%%%%%%%%%%%%%%%%%%%%%%%%%%%%%%%%%%%%%%%%%%%%%%%%%%%%%%
\subsection{\label{ssec:5c}Examples}

We revisit the three examples introduced in Section \ref{ssec:4c} and compute the various quantities in Theorems \ref{thm:4} and \ref{thm:5}.

{\bf Example \ref{exa:1}: Lognormal shadowing (continued).} Fix $D_i = r$ for each $i$ and let $\Psi$ be lognormal distributed with parameter $\sigma^2$, i.e., $\Psi \sim \mathcal{LN}(0,\sigma^2)$, where $f_{\Psi}(\psi)$ is given by (\ref{eqn:am}).  Under threshold based transmission decisions, the distribution $F_{W|t}$ in (\ref{eqn:be}) is given by
\begin{equation}
F_{W|t}(w) = 1 - \frac{\bar{F}_{\Psi}(w r^{\alpha})}{\bar{F}_{\Psi}(t r^{\alpha})} = 1 - \frac{Q\left( \frac{\log (w r^{\alpha})}{\sigma} \right)}{Q\left( \frac{\log(t r^{\alpha})}{\sigma} \right)},
\end{equation}
where $Q(\cdot)$ is the CCDF of the $\mathcal{N}(0,1)$ standard normal distribution.  It is straightforward to establish that:
\begin{equation}
\mathbb{E}[\Psi^{-\delta} \mathbf{1}_{\Psi > t r^{\alpha}}] = e^{\frac{(\delta \sigma)^2}{2}} Q\left(\frac{\log (t r^{\alpha})}{\sigma} + \delta \sigma \right), \label{eqn:bo}
\end{equation}
It follows that $\kappa(t)$ in (\ref{eqn:bh}) is
\begin{equation}
\label{eqn:bp}
\kappa(t) = \pi e^{\delta^2 \sigma^2} r^2 \frac{Q\left(\frac{\log (t r^{\alpha})}{\sigma} + \delta \sigma \right)}{Q\left(\frac{\log (t r^{\alpha})}{\sigma}\right)}.
\end{equation}
Moreover, $\gamma(t)$ in Theorem \ref{thm:5} is given by
\begin{equation}
\label{eqn:bq}
\gamma(t) = \pi r^2 \lambda \beta^{\delta} e^{\delta^2 \sigma^2} Q\left(\frac{\log(t r^{\alpha})}{\sigma} + \delta \sigma \right).
\end{equation}
Solving $\gamma(t) = g$ for $t$ yields
\begin{equation}
\label{eqn:br}
\gamma^{-1}(g) = r^{-\alpha} \exp \left\{\sigma Q^{-1} \left( \frac{g}{\pi r^2 \lambda \beta^{\delta} e^{\delta^2 \sigma^2}} \right) - \delta \sigma^2  \right\}.
\end{equation}

{\bf Example \ref{exa:2}: Rayleigh fading (continued).} Fix $D_i = r$ for each $i$ and let $\Psi \sim {\rm Exp}(1)$.  Under threshold based transmission decisions, the distribution $F_{W|t}$ in (\ref{eqn:be}) is given by
\begin{equation}
\label{eqn:bs}
F_{W|t}(w) = 1 - \exp \{ -(w-t) r^{\alpha}\}, ~ w > t.
\end{equation}
It is straightforward to establish that:
\begin{equation}
\label{eqn:bt}
\mathbb{E}[\Psi^{-\delta} \mathbf{1}_{\Psi > t r^{\alpha}}] = \Gamma(1-\delta,t r^{\alpha}),
\end{equation}
where $\Gamma(a,x)$ is the incomplete Gamma function.  It follows that $\kappa(t)$ in (\ref{eqn:bh}) is
\begin{equation}
\label{eqn:bu}
\kappa(t) = \pi \Gamma(1+\delta) \Gamma(1-\delta,t r^{\alpha}) e^{t r^{\alpha}} r^2.
\end{equation}
Substituting $t = 0$ is consistent with (\ref{eqn:au}) since $\lim_{x \to 0} \Gamma(1-\delta,x) e^x = \Gamma(1-\delta)$.  Moreover, $\gamma(t)$ in Theorem \ref{thm:5} is given by
\begin{equation}
\label{eqn:bv}
\gamma(t) = \pi r^2 \lambda \beta^{\delta} \Gamma(1+\delta) \Gamma(1-\delta,t r^{\alpha}).
\end{equation}
Solving $\gamma(t) = g$ for $t$ yields
\begin{equation}
\label{eqn:bw}
\gamma^{-1}(g) = r^{-\alpha} \Gamma^{-1}\left(1-\delta,\frac{g}{\pi r^2 \lambda \beta^{\delta} \Gamma(1+\delta)} \right),
\end{equation}
where $\Gamma^{-1}(a,y)$ solves $\Gamma(a,\Gamma^{-1}(a,y)) = y$.

{\bf Example \ref{exa:3}: Nearest receiver transmissions (continued).} Fix $\Psi_{ij} = \psi$ for each $i,j$, and let $D_i \sim F_D$ in (\ref{eqn:aw}). Under threshold based transmission decisions, the distribution $F_{W|t}$ in (\ref{eqn:be}) is given by
\begin{eqnarray}
F_{W|t}(w) &=& 1 - \frac{F_D((\psi/w)^{\frac{1}{\alpha}})}{F_D((\psi/t)^{\frac{1}{\alpha}})} \nonumber \\
&=& 1 - \frac{1-\exp\{-\pi \lambda' (\psi/w)^{\delta}\}}{1-\exp\{-\pi \lambda' (\psi/t)^{\delta}\}}. \label{eqn:bx}
\end{eqnarray}
It is straightforward to establish that:
\begin{equation}
\label{eqn:by}
\mathbb{E}[D^2 \mathbf{1}_{D < (\psi/t)^{\frac{1}{\alpha}}}] = \frac{1 - (1 + \pi \lambda' (\psi/t)^{\delta}) e^{-\pi \lambda' (\psi/t)^{\delta}}}{\pi \lambda'}.
\end{equation}
It follows that $\kappa(t)$ in (\ref{eqn:bh}) is
\begin{equation}
\label{eqn:bz}
\kappa(t) = \frac{1}{\lambda'} \cdot \frac{1 - (1 + \pi \lambda' (\psi/t)^{\delta})e^{-\pi \lambda' (\psi/t)^{\delta}}}{1-e^{- \pi \lambda' (\psi/t)^{\delta}}}.
\end{equation}
Moreover, $\gamma(t)$ in Theorem \ref{thm:5} is given by
\begin{equation}
\label{eqn:ca}
\gamma(t) = \frac{\lambda}{\lambda'} \beta^{\delta} \left[1 - (1+ \pi \lambda' (\psi/t)^{\delta}) e^{-\pi \lambda' (\psi/t)^{\delta}} \right].
\end{equation}
Solving $\gamma(t) = g$ for $t$ yields
\begin{equation}
\label{eqn:cb}
\gamma^{-1}(g) = \psi \left[ - \frac{1}{\pi \lambda'} \left(1 + W \left(\frac{-1}{e}\left(1 - \frac{\lambda' g}{\lambda \beta^{\delta}}\right) \right) \right) \right]^{-\frac{\alpha}{2}},
\end{equation}
where $W(x)$ is the $k=-1$ branch of the Lambert $W$ function, satisfying $W(x) e^{W(x)} = x$ \cite{CorGon1996}.

%%%%%%%%%%%%%%%%%%%%%%%%%%%%%%%%%%%%%%%%%%%%%%%%%%%%%%%%%%%%%%%%%%%%%%%
% 6. Numerical and simulation results
%%%%%%%%%%%%%%%%%%%%%%%%%%%%%%%%%%%%%%%%%%%%%%%%%%%%%%%%%%%%%%%%%%%%%%%
\section{\label{sec:6}Numerical and simulation results}

We present numerical and simulation results for the three examples studied in Sections \ref{sec:4} and \ref{sec:5}.  Numerical results are computed using Mathematica, and our simulator is written in Perl.  The simulation methodology is described in greater detail in \cite{WebKam2005}.  Throughout this section we set
\begin{equation}
\label{eqn:cc}
\begin{array}{cccc}
\alpha = 4, & \delta = \frac{1}{2}, & \beta = 3, & \lambda = \frac{1}{100}.
\end{array}
\end{equation}
For Examples 1 and 2 where the distances $\{D_i\} = r$ are constant, we set $r = \frac{1}{2 \sqrt{\lambda}} = 5$ meters.  For Example 3, where the fading coefficients $\{\Psi_{ij}\} = \psi$ are constant, we set $\psi = 1$.  The abscissa in all plots of outage probability ($q$) and spatial throughput ($\tau)$ is $p$, the probability of transmission under the randomized transmission rule.  The abscissa is also labeled $p$ for the plots of $q$ and $\tau$ under the threshold transmission rule with threshold $t$.  This is done to facilitate comparison between the performance plots under the two scheduling rules.  For each $p \in [0,1]$ the threshold $t$ is chosen so that the intensity of points under the two transmission rules is the same, i.e.,
\begin{equation}
\label{eqn:cd}
\mu(p) = \mu(t) ~ \Leftrightarrow ~ \lambda p = \lambda \bar{F}_W(t) ~ \Leftrightarrow ~ t(p) = \bar{F}_W^{-1}(p).
\end{equation}
It follows that the threshold is computed from $p$ using the function $t(p) = \bar{F}_W^{-1}(p)$.

%%%%%%%%%%%%%%%%%%%%%%%%%%%%%%%%%%%%%%%%%%%%%%%%%%%%%%%%%%%%%%%%%%%%%%%
% 6-A. Example 1: Lognormal shadowing
%%%%%%%%%%%%%%%%%%%%%%%%%%%%%%%%%%%%%%%%%%%%%%%%%%%%%%%%%%%%%%%%%%%%%%%
\subsection{\label{ssec:6a}Example 1: Lognormal shadowing}

This section presents numerical and simulation results from Example 1, where the $\{D_i\}$ are constant and the $\{\Psi_{ij}\}$ are lognormal random variables with PDF (\ref{eqn:am}) and $\sigma = \frac{\log 10}{10} 6$ ($6$ dB).  Solving $t(p) = \bar{F}_W^{-1}(p)$ in (\ref{eqn:cd}) yields $t(p) = r^{-\alpha} {\rm exp}(\sigma Q^{-1}(p))$, where $Q^{-1}(\cdot)$ is the inverse of the standard normal CCDF.

Figure \ref{fig:1} shows four plots of simulation and numerical results for lognormal shadowing.  The top left plot illustrates the lower and upper bounds along with simulation results for the outage probability $q$ versus the transmission probability $p$ under randomized and threshold transmissions with no channel inversion.  The bounds are seen to be reasonably tight, especially for the usual case of interest where $q$ is small.  Furthermore, the dramatic reduction in outage achievable through threshold scheduling is apparent.  The other three plots show simulation results for the four cases discussed in the paper: randomized transmissions with and without channel inversion, and threshold transmissions with and without channel inversion.  The three plots are of outage probability $q$ versus transmission probability $p$, spatial throughput $\tau$ versus $p$, and the transmission capacity $c$ versus the outage requirement $\epsilon$.

Several observations merit comment.  First, note that as $p \to 0$ (as $t(p) \to \infty$) channel inversion has no impact on performance, while as $p \to 0$ (as $t(p) \to \infty$) randomized and threshold transmissions have identical performance.  The limited impact of channel inversion is especially apparent for  threshold transmissions: for large $t$ (small $p$) only potential transmitters with good channels elect to transmit, and thus there is less need for using channel inversion to compensate for poor channels compared with smaller $t$ (larger $p$).  For large $p$ (small $t$) employing thresholds is  increasingly ineffective in restricting transmissions to nodes with good channels, and as such becomes equivalent to randomized transmissions.

Second, as discussed in the comments of Section \ref{sec:4} and \ref{sec:5}, channel inversion always reduces performance (larger $q$, smaller $\tau$, smaller $c$).   As discussed, channel inversion is a compensation mechanism allowing nodes with poor channels to their receivers to obtain acceptable received signal power.  Although nodes with good channels reduce their power and hence decrease the amount of interference they cause, the net effect is undesirable.  The intuitive explanation, as noted in Section \ref{ssec:4d}, is that a single dominant interferer is the most likely contributor to causing outage, so a policy of pairwise channel inversion increases the likelihood of a dominant interference event occurring since the chances of at least one nearby transmitter having a poor channel to its receiver are reasonably good.

Third, note that the plots of spatial throughput and transmission capacity achieve the same peaks (for each of the four cases), and that $i)$ the peak capacity under threshold transmissions is over twice the peak capacity under randomized transmissions, and $ii)$ the outage requirement $\epsilon$ required to obtain those peaks is much smaller for threshold transmissions than for randomized transmissions.  In effect, channel variability should not be dealt with through randomization, but instead should be exploited through threshold scheduling.

%%%%%%%%%%%%%%%%%%%%%%%%%%%%%%%%%%%%%%%%%%%%%%%%%%%%%%%%%%%%%%%%%%%%%%%
% 6-B. Example 2: Rayleigh fading
%%%%%%%%%%%%%%%%%%%%%%%%%%%%%%%%%%%%%%%%%%%%%%%%%%%%%%%%%%%%%%%%%%%%%%%
\subsection{\label{ssec:6b}Example 2: Rayleigh fading}

For Example 2 the $\{D_i\}$ are constant and the $\{\Psi_{ij}\}$ are exponentially distributed with parameter $1$.  Solving $t(p) = \bar{F}_W^{-1}(p)$ in (\ref{eqn:cd}) yields $t(p) = -\frac{\log p}{r^{\alpha}}$. Figure \ref{fig:2} shows four plots of simulation and numerical results for Rayleigh fading.  The top left plot illustrates the upper and lower bounds and the simulation results spatial throughput $\tau$ versus the transmission probability $p$ for threshold based transmissions with channel inversion (dashed curves) and without channel inversion (solid curves). The other three plots are for the same scenarios as in Fig. \ref{fig:1}.  The comments made in Example 1 hold here as well, but overall the effects are not quite as severe as the channel variations are not quite as large as in the lognormal case with $\sigma = 6$ dB.

%%%%%%%%%%%%%%%%%%%%%%%%%%%%%%%%%%%%%%%%%%%%%%%%%%%%%%%%%%%%%%%%%%%%%%%
% 6-C. Example 3: Nearest receiver transmissions
%%%%%%%%%%%%%%%%%%%%%%%%%%%%%%%%%%%%%%%%%%%%%%%%%%%%%%%%%%%%%%%%%%%%%%%
\subsection{\label{ssec:6c}Example 3: Nearest receiver transmissions}

In Example 3 the $\{D_i\}$ have CCDF (\ref{eqn:aw}) and the $\{\Psi_{ij}\}$ are constant.  Solving $t(p) = \bar{F}_W^{-1}(p)$ yields $t(p) = \psi (\frac{\pi \lambda'}{-\log(1-p)})^{\frac{1}{\delta}}$.  The density $\lambda'$ of potential receivers in $\Pi'$ is set equal to the density of potential transmitters: $\lambda' = \lambda$.

Figure \ref{fig:3} shows four plots of simulation and numerical results for nearest receiver transmissions.  The top left plot illustrates the transmission capacity $c$ versus the outage requirement $\epsilon$ for randomized transmissions (solid curves) and threshold transmissions (dashed curves) with channel inversion.  The three curves for each case are the lower and upper bounds along with simulation results.  The bounds are again seen to be reasonably tight, especially for the small $\epsilon$ case of usual interest.  The other three plots follow as in Examples 1 and 2.

The comments from Examples 1 and 2 apply here as well; in other words random hop distances behave much like random channel effects. Note that the bottom right plot of transmission capacity illustrates that threshold scheduling achieves a three fold or greater increase in transmission capacity relative to the capacity under randomized transmissions for all $\epsilon < 0.2$. Even more than Examples 1 and 2, this example highlights the tension between throughput and fairness: in this example performance is maximized by employing a threshold rule such that only nodes that are sufficiently close to their intended receivers are selected to transmit. In Example 2 and to a lesser extent Example 1, one could argue that the channel coherence timescale is short enough such that the unfairness of opportunistic scheduling is of limited concern to the typical user.  Here, however, the transmitter receiver distances are changing on the timescale of user mobility, meaning the unfairness of opportunistic scheduling is of much greater concern.

%%%%%%%%%%%%%%%%%%%%%%%%%%%%%%%%%%%%%%%%%%%%%%%%%%%%%%%%%%%%%%%%%%%%%%%
% 7. Conclusion
%%%%%%%%%%%%%%%%%%%%%%%%%%%%%%%%%%%%%%%%%%%%%%%%%%%%%%%%%%%%%%%%%%%%%%%
\section{\label{sec:7}Conclusion}

The goal of this paper was to develop a methodology, and some insights, on how ad hoc network capacity is affected by temporal variations in channel quality and transmission distance.  We focused on the case where each node has only local information; in particular it knows the channel to a desired receiver.  This approach, while suboptimal, has the considerable merits of being realistic and analytically tractable.

We made the following observations, with transmission capacity and outage probability as the metrics of interest. First, randomized transmissions perform poorly in the presence of either fading or variable channel distances. That is, variability strictly reduces capacity in transmitters that are blind to the channel.  Second, we showed that a policy of channel inversion, while helping users with poor channels, negatively impacts the overall network capacity.  The intuition is that this increases the likelihood of a dominant interferer causing an outage: because of the coupled nature of all the links in an ad hoc network, channel inversion causes all the nodes in an area to suffer when a single link is poor.  Third, channel variations can be exploited through the use of simple threshold scheduling, where a user transmits when its desired channel is above a target.  We derive the optimal threshold, and show that over many ranges of interest, the capacity is about three times higher than with no scheduling.

%\bibliographystyle{IEEEtran}

%%%%%%%%%%%%%%%%%%%%%%%%%%%%%%%%%%%%%%%%%%%%%%%%%%%%%%%%%%%%%%%%%%%%%%%
% A. Appendix
%%%%%%%%%%%%%%%%%%%%%%%%%%%%%%%%%%%%%%%%%%%%%%%%%%%%%%%%%%%%%%%%%%%%%%%

\appendix

%%%%%%%%%%%%%%%%%%%%%%%%%%%%%%%%%%%%%%%%%%%%%%%%%%%%%%%%%%%%%%%%%%%%%%%
% A-1. Proof of Theorem 2
%%%%%%%%%%%%%%%%%%%%%%%%%%%%%%%%%%%%%%%%%%%%%%%%%%%%%%%%%%%%%%%%%%%%%%%
\section*{Proof of Theorem \ref{thm:2}}

The proof consists of three steps: $i)$ obtaining the lower bound ($\bar{F}_Y^l(y)$), $ii)$ obtaining the upper bound ($\bar{F}_Y^u(y)$), and $iii)$ obtaining the asymptotic expansions for $\bar{F}_Y(y), \bar{F}_Y^l(y)$, and $\bar{F}_Y^u(y)$.  We begin with some definitions.

Under randomized transmissions and without channel inversion, the normalized aggregate interference seen by the reference receiver, $Y$, is given by (\ref{eqn:m}).  It is clear that $Y$ is the product of a r.v. $1/W_0$ and a shot noise process with points $\{X_i\}$ and marks $\{\Psi_{i0}\}$.  The $\{\Psi_{i0}\},\{X_i\}$ and $W_0$ are all mutually independent.  Fix the outage threshold at $y > 0$ and the received signal power at $W_0 = w$.  Split $\Phi$ into two disjoint complementary processes: $\Phi = \Phi_{y,w} \cup \Phi_{y,w}^c$, where:
\begin{eqnarray}
\Phi_{y,w} &=& \left\{X_i \in \Phi :  \frac{1}{w} \Psi_{i0} |X_i|^{-\alpha} \geq y \right\}, \nonumber \\
\Phi_{y,w}^c &=& \left\{X_i \in \Phi : \frac{1}{w} \Psi_{i0} |X_i|^{-\alpha} < y \right\}. \label{app:a}
\end{eqnarray}
Thus $\Phi_{y,w}$ is the set of points that are individually capable of causing outage at the reference receiver if the outage threshold is $y$ and the received signal power is $w$.  It is helpful to think of the points in $\Phi_{y,w}$ as the dominant interferers for the reference receiver, and the remaining points in $\Phi_{y,w}^c$ as the non-dominant interferers.  Note that although the quantities $\Psi_{i0}$ and $X_i$ are independent for each $i$ in $\Phi$, they are not independent in $\Phi_{y,w}$ and $\Phi_{y,w}^c$.  Also, although $\Phi$ is a {\em stationary} (homogeneous) Poisson process of intensity $\mu$, both $\Phi_{y,w}$ and $\Phi_{y,w}^c$ are non-stationary Poisson processes.  Define the aggregate normalized interference from these processes as
\begin{equation}
\label{app:b}
Y_{y,w} = \sum_{i \in \Phi_{y,w}} \Psi_{i0} |X_i|^{-\alpha}, ~~ Y_{y,w}^c = \sum_{i \in \Phi_{y,w}^c} \Psi_{i0} |X_i|^{-\alpha},
\end{equation}
and note that $Y = Y_{y,w} + Y_{y,w}^c$.

{\bf Step 1: lower bound $\bar{F}_Y^l(y)$.}
The lower bound on $\bar{F}_Y(y)$ is
\begin{eqnarray}
\bar{F}_Y(y) & \geq & \bar{F}_Y^l(y) = \mathbb{P}(Y_{y,W_0} > y) \nonumber \\
&=& \int_0^{\infty} \mathbb{P}(Y_{y,w} > y) {\rm d}F_W(w). \label{app:c}
\end{eqnarray}
To compute the lower bound observe that the event $\{Y_{y,w} > y\}$ is the same as the event $\{\Phi_{y,w} \neq \emptyset\}$.  With this observation we can compute the lower bound using the expression for the void probability of a Poisson process:
\begin{eqnarray}
\mathbb{P}(Y_{y,w} > y) &=& 1 - \mathbb{P}(\Phi_{y,w} = \emptyset) \nonumber \\
&=& 1 - \exp \left\{ -\int_{\mathbb{R}^2} \mu_{y,w}(x) {\rm d}x \right\}, \label{app:d}
\end{eqnarray}
where $\mu_{y,w}(x)$ is the density of points in $\Phi_{y,w}$ at location $x$:
\begin{equation}
\label{app:e}
\mu_{y,w}(x) = \mu \mathbb{P}\left(\frac{1}{w} \Psi |x|^{-\alpha} \geq y\right).
\end{equation}
Noting that the density is radially symmetric, we can switch to polar coordinates, with slight abuse of notation writing $\mu_{y,w}(r)$ for the intensity of $\Phi_{y,w}$ at distance $r$.  The resulting expression is simplified by writing $\bar{F}_{\Psi}(s) = \int_s^{\infty} f_{\Psi}(\psi) {\rm d}\psi$, exchanging the order of integration, and using the change of variables $s = y r^{\alpha}$:
\begin{eqnarray}
\bar{F}_Y^l(y) &=& 1 - \int_0^{\infty} e^{ -2 \pi \mu \int_0^{\infty} \mathbb{P}(\Psi > w y r^{\alpha}) r {\rm d}r } {\rm d}F_W(w) \nonumber \\
&=& 1 - \mathbb{E} \left[ \exp \left\{ - \pi \mu \mathbb{E}[\Psi^{\delta}] W^{-\delta} y^{-\delta} \right\} \right] \nonumber \\
&=& 1 - \mathbb{E} \left[ \exp \left\{ - K \mu  y^{-\delta} \right\} \right], \label{app:f}
\end{eqnarray}
where $K = \pi \mathbb{E}[\Psi^{\delta}] \Psi^{-\delta} D^2$.  This completes the lower bound.

{\bf Step 2: upper bound $\bar{F}_Y^u(y)$.}
To establish the upper bound, we condition on $Y_{y,w}$ for each $w$:
\begin{eqnarray}
\bar{F}_Y(y) &=& \int_0^{\infty} {\rm d}F_W(w) \left[ \mathbb{P}(Y > y | Y_{y,w} > y) \bar{F}_{Y_{y,w}}(y) + \right. \nonumber \\
& & \left. \mathbb{P}(Y > y | Y_{y,w} \leq y) F_{Y_{y,w}}(y) \right]  \nonumber \\
&=& \int_0^{\infty} {\rm d}F_W(w) \left[ 1 \cdot \bar{F}_{Y_{y,w}}(y) + \bar{F}_{Y_{y,w}^c}(y) F_{Y_{y,w}}(y) \right]  \nonumber \\
&=& \int_0^{\infty} {\rm d}F_W(w) \left[ \left(1 - \exp \{- \pi \mu \mathbb{E}[\Psi^{\delta}] w^{-\delta} y^{-\delta}\} \right)  + \right. \nonumber \\
& & \left. \bar{F}_{Y_{y,w}^c}(y) \exp \{- \pi \mu \mathbb{E}[\Psi^{\delta}] w^{-\delta} y^{-\delta}\} \right]  \nonumber \\
&=& 1 - \int_0^{\infty} {\rm d}F_W(w) (1 - \bar{F}_{Y_{y,w}^c}(y)) \times \nonumber \\
& & \exp \{- \pi \mu \mathbb{E}[\Psi^{\delta}] w^{-\delta} y^{-\delta}\}  \nonumber \\
&=& 1 - \mathbb{E} \left[ (1 - \bar{F}_{Y_{y,W}^c}(y)) e^{-\pi \mu \mathbb{E}[\Psi^{\delta}] W^{-\delta} y^{-\delta}} \right], \label{app:g}
\end{eqnarray}
where we have used the fact that $F_{Y_{y,w}}(y) = \exp \{- \pi \mu \mathbb{E}[\Psi^{\delta}] w^{-\delta} y^{-\delta}\}$.  The upper bound $\bar{F}_Y^u(y)$ on $\bar{F}_Y(y)$ is obtained by finding an upper bound $\bar{F}_{Y_{y,w}^c}^u(y)$ on $\bar{F}_{Y_{y,w}^c}(y)$ for each $w$.  Application of the Chebychev inequality yields:
\begin{equation}
\bar{F}_{Y_{y,w}^c}(y) \leq \bar{F}_{Y_{y,w}^c}^u(y) = \frac{{\rm Var}(Y_{y,w}^c)}{(y - \mathbb{E}[Y_{y,w}^c])^2}. \label{app:h}
\end{equation}
We apply Campbell's Theorem \cite{StoKen1996} to compute the mean and variance of $Y_{y,w}^c$.  This requires we characterize the density of points in $\Phi_{y,w}^c$ in the product space $\mathbb{R}^2 \times \boldsymbol{\Psi}$, where $\boldsymbol{\Psi} = \mathbb{R}^+$ is the support of $\Psi$.  The density at a point $(x,\psi)$ in this space is $\mu_{y,w}^c(x,\psi) = \mu f_{\Psi}(\psi) \mathbf{1}_{|x|^{-\alpha} \psi < w y}$.  Straightforward analysis yields:
\begin{eqnarray}
\mathbb{E}[Y_{y,w}^c]
&=& \int_{\mathbb{R}^2} \int_0^{\infty} \frac{\psi}{w} |x|^{-\alpha} \mu_{y,w}^c(x,\psi) {\rm d}x {\rm d}\psi  \nonumber \\
&=& 2 \pi \mu \int_0^{\infty} \frac{\psi}{w} \left( \int_{(\frac{\psi}{wy})^{\frac{1}{\alpha}}}^{\infty} r^{1-\alpha} {\rm d}r \right) {\rm d}F_{\Psi}(\psi) \nonumber \\
&=& \frac{2 \pi \mu}{\alpha-2} y^{1-\frac{2}{\alpha}} w^{-\frac{2}{\alpha}}\int_0^{\infty} \psi^{\frac{2}{\alpha}} {\rm d}F_{\Psi}(\psi) \nonumber \\
&=& \frac{\delta}{1-\delta} \pi \mu \mathbb{E}[\Psi^{\delta}] w^{-\delta} y^{1-\delta}. \label{app:i}
\end{eqnarray}
Similarly,
\begin{eqnarray}
{\rm Var}(Y_{y,w}^c)
&=& \int_{\mathbb{R}^2} \int_0^{\infty} \left( \frac{\psi}{w} \right)^2 |x|^{-2\alpha} \mu_{y,w}^c(x,\psi) {\rm d}x {\rm d}\psi \nonumber \\
&=& 2 \pi \mu \int_0^{\infty} \left( \frac{\psi}{w} \right)^2 \left( \int_{(\frac{\psi}{wy})^{\frac{1}{\alpha}}}^{\infty} r^{1-2\alpha} {\rm d}r \right) {\rm d}F_{\Psi}(\psi) \nonumber \\
&=& \frac{2 \pi \mu}{2\alpha-2} y^{2-\frac{2}{\alpha}} \int_0^{\infty} \left( \frac{\psi}{w} \right)^{\frac{2}{\alpha}} {\rm d}F_{\Psi}(\psi) \nonumber \\
&=& \frac{\delta}{2-\delta} \pi \mu \mathbb{E}[\Psi^{\delta}] w^{-\delta} y^{2-\delta}. \label{app:j}
\end{eqnarray}
Using $K = \pi \mathbb{E}[\Psi^{\delta}] \Psi^{-\delta} D^2$ it follows that the upper bound may be expressed as
\begin{equation}
\label{app:k}
\bar{F}_Y(y) \leq 1 - \mathbb{E}\left[\left(1 - \frac{\frac{\delta}{2-\delta}K \mu y^{-\delta}}{(1-\frac{\delta}{1-\delta} K \mu y^{-\delta})^2} \right)^+ e^{-K \mu y^{-\delta}} \right],
\end{equation}
where $(x)^+ = \max\{x,0\}$.

{\bf Step 3: asymptotic expansions.}
We next obtain the asymptotic expansions of $\bar{F}_Y(y),\bar{F}_Y^l(y),\bar{F}_Y^u(y)$ as $y \to \infty$.  In all three cases we will compute the series representations of the conditional distributions $\bar{F}_{Y|W}(y|w),\bar{F}_{Y|W}^l(y|w),\bar{F}_{Y|W}^u(y|w)$ conditioned on $W_0 = w$, then recover the unconditioned distributions by integrating against the distribution $F_W$.  We first obtain the series expansion of $\bar{F}_Y(y)$.  Equation (29) in \cite{LowTei1990} gives the series representation of the PDF of $I(t)$ in (\ref{eqn:d}) when $h(t)$ is a power law (\ref{eqn:e}) with marks $\{K_i\}$ and $A = 0, B = \infty$ and $\beta > 1$:
\begin{eqnarray}
f_{I(t)}(x) &=&  \frac{1}{\pi x} \sum_{n=1}^{\infty} \frac{(-1)^{n+1}}{n!} \Gamma(1+n/\beta) \sin(\pi n/\beta) ~ \times \nonumber \\
& & \left[\mu_1 \Gamma(1-1/\beta) \mathbb{E}[K^{\frac{1}{\beta}}] x^{-\frac{1}{\beta}} \right]^n. \label{app:l}
\end{eqnarray}
Here $\mu_1$ is the intensity of the Poisson process of times $\{t_j\}$ in (\ref{eqn:d}).  Recall that $I(t)$ is a shot noise process on $\mathbb{R}$, not $\mathbb{R}^2$.  We can nonetheless use the result to obtain the series representation of the CCDF of $Y$ by translating the shot noise process on $\mathbb{R}^2$ onto $\mathbb{R}$, then integrating the PDF to get the CCDF.  Translating a Poisson shot noise point process on $\mathbb{R}^2$ onto $\mathbb{R}$ is discussed in \cite{IloHat1998} and \cite{Hae2005}; in essence, the path loss exponent changes from $\alpha$ to $\alpha/2$ and the intensity increases from $\lambda$ to $\pi \lambda$.  Applying this transformation yields the conditional PDF of $Y$:
\begin{eqnarray}
f_{Y|W}(y|w) &=& \frac{1}{\pi w y} \sum_{n=1}^{\infty} \frac{(-1)^{n+1}}{n!} \Gamma(1+n \delta) \sin(\pi n \delta) ~ \times \nonumber \\
& & \left[ \pi \mu \Gamma(1-\delta) \mathbb{E}[\Psi^{\delta}] (wy)^{-\delta} \right]^n. \label{app:m}
\end{eqnarray}
Integrating the conditional PDF yields the conditional CCDF
\begin{eqnarray}
\bar{F}_{Y|W}(y|w) &=& \sum_{n=1}^{\infty} \frac{(-1)^{n+1}}{n!} \frac{1}{\pi n \delta} \Gamma(1+n \delta) \sin(\pi n \delta) ~ \times \nonumber \\
& & \left[ \pi \mu \Gamma(1-\delta) \mathbb{E}[\Psi^{\delta}] (wy)^{-\delta} \right]^n. \label{app:n}
\end{eqnarray}
Taking the dominant $n=1$ term of the series yields
\begin{equation}
\label{app:o}
\bar{F}_{Y|W}(y|w) = \mu \pi \mathbb{E}[\Psi^{\delta}] (wy)^{-\delta} + O((wy)^{-2\delta}),
\end{equation}
where we have used the identity $\frac{\sin(\pi \delta)}{\pi \delta}\Gamma(1+\delta) \Gamma(1-\delta) = 1$.  Unconditioning yields
\begin{eqnarray}
\bar{F}_Y(y)
&=& \int_0^{\infty} \bar{F}_{Y|W}(y|w) {\rm d}F_W(w) \nonumber \\
&=& \mu \pi \mathbb{E}[\Psi^{\delta}] \mathbb{E}[W^{-\delta}] y^{-\delta} + O(y^{-2\delta}). \label{app:p}
\end{eqnarray}

We next obtain the series representation of the conditional lower bound $\bar{F}_{Y|W}^l(y|w)$:
\begin{eqnarray}
\bar{F}_{Y|W}^l(y|w) &=& 1-\exp \left\{ -\mu \pi \mathbb{E}[\Psi^{\delta}] (wy)^{-\delta} \right\} \nonumber \\
&=& 1 - \left(1 + \sum_{n=1}^{\infty} \frac{(-1)^{n} ( \mu \pi \mathbb{E}[\Psi^{\delta}] (wy)^{-\delta})^n}{n!} \right) \nonumber \\
&=& \sum_{n=1}^{\infty} \frac{(-1)^{n+1}}{n!} (\mu \pi \mathbb{E}[\Psi^{\delta}])^n (wy)^{-n \delta}. \label{app:q}
\end{eqnarray}
Taking the dominant $n=1$ term of the series yields
\begin{equation}
\label{app:r}
\bar{F}_{Y|W}^l(y|w) = \mu \pi \mathbb{E}[\Psi^{\delta}] (wy)^{-\delta} + O((wy)^{-2\delta}),
\end{equation}
and unconditioning yields
\begin{equation}
\label{app:s}
\bar{F}_Y^l(y) = \mu \pi \mathbb{E}[\Psi^{\delta}] \mathbb{E}[W^{-\delta}] y^{-\delta} + O(y^{-2\delta}).
\end{equation}

Finally, we obtain the first order Taylor series expansion of the conditional upper bound $\bar{F}_{Y|W}^u(y|w)$.  Define $x = \mu \pi \mathbb{E}[\Psi^{\delta}] (wy)^{-\delta}$ so that the conditional lower bound is given by:
\begin{equation}
\label{app:t}
g(x) = 1 - \left(1 - \frac{\frac{\delta}{2-\delta} x}{(1 - \frac{\delta}{1-\delta} x)^2} \right)^+ e^{-x}.
\end{equation}
Taking an expansion around $x = 0$ corresponds to finding the asymptotic order for large $y$.  The first order Taylor series expansion of $g(x)$ around $x = 0$ is easily seen to be
\begin{equation}
\label{app:u}
g(x) = g(0) + g'(0) x + O(x^2) = 0 + \frac{2}{2-\delta} x + O(x^2).
\end{equation}
Substituting back $\mu \pi \mathbb{E}[\Psi^{\delta}] (wy)^{-\delta}$ for $x$ and unconditioning yields
\begin{equation}
\label{app:v}
\bar{F}_Y^u(y) = \frac{2}{2-\delta} \mu \pi \mathbb{E}[\Psi^{\delta}] \mathbb{E}[W^{-\delta}] y^{-\delta} + O(y^{-2\delta}).~  \blacksquare
\end{equation}

%%%%%%%%%%%%%%%%%%%%%%%%%%%%%%%%%%%%%%%%%%%%%%%%%%%%%%%%%%%%%%%%%%%%%%%
% A-2. Proof of Corollary 2
%%%%%%%%%%%%%%%%%%%%%%%%%%%%%%%%%%%%%%%%%%%%%%%%%%%%%%%%%%%%%%%%%%%%%%%
\section*{Proof of Corollary \ref{cor:2}}

The corollary follows from the proof of Theorem \ref{thm:2} with the following changes.  Note that under randomized transmissions with channel inversion, the normalized aggregate interference seen at the reference receiver, $Y$, given by (\ref{eqn:aa}), is a shot noise process, with random marks $Z_i = \frac{\Psi_{i0}}{W_{ii}}$.  There is no need to condition on the received signal power.  This means that $\Phi_{y,w}$ and $\Phi_{y,w}^c$ in (\ref{app:a}) may be replaced with
\begin{eqnarray}
\Phi_y &=& \left\{X_i \in \Phi :  Z_i |X_i|^{-\alpha} \geq y \right\} \nonumber \\
\Phi_y^c &=& \left\{X_i \in \Phi : Z_i |X_i|^{-\alpha} < y \right\}, \label{app:w}
\end{eqnarray}
and $Y_{y,w}, Y_{y,w}^c$ in (\ref{app:b}) may be replaced with
\begin{equation}
\label{app:x}
Y_y = \sum_{i \in \Phi_y} Z_i |X_i|^{-\alpha}, ~~ Y_y^c = \sum_{i \in \Phi_y^c} Z_i |X_i|^{-\alpha}.
\end{equation}

{\bf Step 1: lower bound $\bar{F}_Y^l(y)$.}
The lower bound (\ref{app:f}) becomes
\begin{eqnarray}
\bar{F}_Y^l(y) &=& 1 - \exp\{ -2 \pi \mu \int_0^{\infty} \mathbb{P}(Z > y r^{\alpha}) r {\rm d}r \} \nonumber \\
&=& 1 - \exp \left\{ - \pi \mu \mathbb{E}[\Psi^{\delta}] \mathbb{E}[W^{-\delta}] y^{-\delta} \right\}  \nonumber \\
&=& 1 - \exp \left\{ - \kappa \mu  y^{-\delta} \right\}. \label{app:y}
\end{eqnarray}

{\bf Step 2: upper bound $\bar{F}_Y^u(y)$.}
The unconditioned version of (\ref{app:g}) is
\begin{eqnarray}
\bar{F}_Y(y) &=& \mathbb{P}(Y > y | Y_y > y) \bar{F}_{Y_y}(y) + \nonumber \\
& & \mathbb{P}(Y > y | Y_y \leq y) F_{Y_y}(y) \nonumber \\
&=& 1 - (1 - \bar{F}_{Y_y^c}(y)) \times \nonumber \\
& & \exp \{- \pi \mu \mathbb{E}[\Psi^{\delta}] \mathbb{E}[W^{-\delta}] y^{-\delta}\}. \label{app:z}
\end{eqnarray}
The upper bound is obtained by applying the Chebychev inequality, replacing (\ref{app:h}) with
\begin{equation}
\bar{F}_{Y_y^c}(y) \leq \bar{F}_{Y_y^c}^u(y) = \frac{{\rm Var}(Y_y^c)}{(y - \mathbb{E}[Y_y^c])^2}. \label{app:aa}
\end{equation}
The mean and the variance become
\begin{eqnarray}
\mathbb{E}[Y_y^c]
&=& \frac{\delta}{1-\delta} \pi \mu \mathbb{E}[\Psi^{\delta}] \mathbb{E}[W^{-\delta}] y^{1-\delta}, \nonumber \\
{\rm Var}(Y_y^c) &=& \frac{\delta}{2-\delta} \pi \mu \mathbb{E}[\Psi^{\delta}] \mathbb{E}[W^{-\delta}] y^{2-\delta}. \label{app:ab}
\end{eqnarray}
Instead of (\ref{app:k}), the unconditioned upper bound is
\begin{equation}
\label{app:ac}
\bar{F}_Y(y) \leq 1 - \left(1 - \frac{\frac{\delta}{2-\delta}\kappa \mu y^{-\delta}}{(1-\frac{\delta}{1-\delta} \kappa \mu y^{-\delta})^2} \right)^+ e^{-\kappa \mu y^{-\delta}}.
\end{equation}
The upper bound is non-trivial when $\bar{F}^u_Y(y) < 1$, where the critical point $\bar{F}^u_Y(y) = 1$ corresponds to the solution of
\begin{equation}
\frac{\delta}{2-\delta} x = \left(1 -  \frac{\delta}{1-\delta} x \right)^2,
\end{equation}
for $x = \kappa \mu y^{-\delta}$.  Solving this equation for $x$ yields the function $h(\delta)$ given in (\ref{eqn:ac}).

{\bf Step 3: asymptotic expansions.}  The asymptotic expansions under channel inversion are the conditional asymptotic expansions from Step 3 of the proof of Theorem \ref{thm:2} with $y$ replacing $wy$ and $Z$ replacing $\Psi$.  $\blacksquare$

%%%%%%%%%%%%%%%%%%%%%%%%%%%%%%%%%%%%%%%%%%%%%%%%%%%%%%%%%%%%%%%%%%%%%%%
% A-3. Chernoff upper bound
%%%%%%%%%%%%%%%%%%%%%%%%%%%%%%%%%%%%%%%%%%%%%%%%%%%%%%%%%%%%%%%%%%%%%%%
\section*{Chernoff upper bound}

Let $\{Z_i\}$, $y$, $\Phi_y$, $\Phi_y^c$, $Y_y$, $Y_y^c$ be as in the proof of Corollary \ref{cor:2}.  In Corollary \ref{cor:2} the Chebychev inequality is used to upper bound $\bar{F}_{Y_y^c}(y)$, which in turn yields an upper bound on the CCDF, $\bar{F}_Y^u(y)$.  Our purpose in this section is to discuss the use of the Chernoff bound instead of the Chebychev bound.  In particular, the Chernoff bound may be used to upper bound $\bar{F}_{Y_y^c}(y)$, which yields a tighter upper bound on the CCDF, $\bar{F}_Y^u(y)$.  The Chernoff rate function requires the log moment generating function (MGF) of the random variable $Y_y^c$, defined as $g_{Y_y^c}(\theta) = \log \mathbb{E}[\exp \{ \theta Y_y^c\}]$.  The log MGF for a functional $\sum_{i \in \Pi} f(x_i,m_i)$ of a non-stationary MPPP $\Pi = \{(X_i,\Psi_i)\}$ with $(x_i,\psi_i) \in \mathcal{S} \times \boldsymbol{\Psi}$ and intensity $\mu(x,\psi)$ is given by Kingman (\cite{Kin1993}, (5.10), page 58) as:
\begin{equation}
\label{app:ad}
g(\theta) = \int_{\mathcal{S}} \int_{\mathcal{\Psi}} \left( e^{\theta f(x,\psi)} - 1 \right) \mu(x,\psi) {\rm d}\psi ~{\rm d}x .
\end{equation}
It is clear that $Y_y^c$ is a functional of a non-stationary MPPP with iid marks $\{Z_i\}$.  The intensity of points from the process $\Phi_y^c$ at the point $(x,z) \in \mathbb{R}^2 \times \mathcal{Z}$ is
\begin{equation}
\label{app:ae}
\mu_{Y_y^c}(x,z) = \mu \frac{{\rm d}}{{\rm d}z} \mathbb{P}(Z |x|^{-\alpha} < y) = \mu f_Z(z) \mathbf{1}_{z < y |x|^{\alpha}}.
\end{equation}
After changing to radial coordinates, it follows that the log MGF for $Y_y^c$ is
\begin{equation}
\label{app:af}
g_{Y_y^c}(\theta) = \mu \int_0^{\infty} \int_0^{y r^{\alpha}} r \left( e^{\theta z r^{-\alpha}} - 1 \right)  {\rm d}F_Z(z) ~  {\rm d}r .
\end{equation}
The Chernoff bound on $\bar{F}_{Y_y^c}(y)$ has a rate function given by the Legendre transform of the log MGF:
\begin{equation}
\label{app:ag}
\bar{F}_{Y_y^c}(y) \leq \exp \left\{ - \sup_{\theta \geq 0} \left[\theta y - g_{Y_y^c}(\theta) \right] \right\}.
\end{equation}
Evaluating the log MGF requires the PDF of the marks $f_Z(z)$, which in turn depends upon the transmission decision policy.  Under randomized transmission decisions with channel inversion the PDF is
\begin{equation}
\label{app:ah}
f_Z(z) = \int_0^{\infty} w f_{\Psi}(wz) {\rm d}F_W(w),
\end{equation}
where $f_W(w)$ is the {\em unconditioned} signal power PDF given by
$\frac{{\rm d}}{{\rm d}w} F_W(w) = \frac{{\rm d}}{{\rm d}w} \mathbb{P}(\Psi D^{-\alpha} < w)$.

Although the Chernoff bound is in principle computable using the above equations, it is in practice often not computationally feasible to do so.  Note that evaluating the PDFs $f_Z(z)$ and $f_{Z|t}(z)$ requires evaluating a double integral, and evaluating the MGF $g_{Y_y^c}(\theta)$ requires evaluating a double integral expressed in terms of $f_Z(z)$, in effect requiring a quadruple integral be evaluated for each $\theta$.  Further, the optimal $\theta$ in the Chernoff rate function must be computed numerically.  In contrast, the Chebychev inequality, although not as tight as the Chernoff bound, is given explicitly without requiring the evaluation of any integrals.   Thus, for both ease of computability and clarity of exposition we have chosen to express our results using the Chebychev bound instead of the Chernoff bound.

%%%%%%%%%%%%%%%%%%%%%%%%%%%%%%%%%%%%%%%%%%%%%%%%%%%%%%%%%%%%%%%%%%%%%%%
% A-4. Proof of Theorem 4
%%%%%%%%%%%%%%%%%%%%%%%%%%%%%%%%%%%%%%%%%%%%%%%%%%%%%%%%%%%%%%%%%%%%%%%
\section*{Proof of Theorem \ref{thm:4}}

Theorem \ref{thm:4} asserts that the sole effect of changing from randomized transmission decisions (without channel inversion) to threshold based transmission decisions (without channel inversion) is to change the distribution of the received signal power.  Consider the expressions for the normalized aggregate interference seen by the reference receiver, $Y$, under randomized transmissions (\ref{eqn:m}) and under threshold transmissions (\ref{eqn:bd}).  They are the same with the exception that $W_0$ in (\ref{eqn:m}) is replaced with $W_{0|t}$ in (\ref{eqn:bd}).  Given that all the results in Theorems \ref{thm:1}, \ref{thm:2}, and \ref{thm:3} are obtained by first conditioning on $W_0$, it follows that those same results will hold under threshold transmissions, with the distribution $F_W$ replaced with $F_{W|t}$.  All that remains is to establish the expression for $\kappa(t)$ in (\ref{eqn:bh}).

Define the random variable $Z = \frac{\Psi}{W_t}$.  The proof consists of developing an expression for $\mathbb{E}[Z^{\delta}]$ in terms of the threshold $t$, the distribution for the channel gains, $F_{\Psi}$, and the distribution for the transmitter to receiver distances, $F_D$.  We first identify the distribution of $W$ conditioned on $W > t$ (step 1), then identify the distribution of $Z$ conditioned on $W > t$ (step 2), and finally compute $\mathbb{E}[Z^{\delta}]$ (step 3).

{\bf Step 1: Distribution of $W$ conditioned on $W > t$.}  The conditioned signal strength distribution may be expressed in terms of the unconditioned signal strength distribution:
\begin{equation}
\label{app:ai}
F_{W|t}(w) = \frac{F_W(w) - F_W(t)}{\bar{F}_W(t)}, ~ w > t,
\end{equation}
with corresponding density
\begin{equation}
\label{app:aj}
f_{W|t}(w) = \frac{f_W(w)}{\bar{F}_W(t)}, ~ w > t.
\end{equation}
The unconditioned signal strength distribution depends upon the distributions
$F_{\Psi},F_D$.  In particular, the unconditioned signal strength CDF is
\begin{equation}
\label{app:ak}
F_W(w) = \int_0^{\infty} F_{\Psi}(w x^{\alpha}) {\rm d}F_D(x),
\end{equation}
and the unconditioned signal strength PDF is
\begin{equation}
\label{app:al}
f_W(w) = \int_0^{\infty} x^{\alpha} f_{\Psi}(w x^{\alpha}) {\rm d}F_D(x).
\end{equation}

{\bf Step 2: Distribution of $Z$ conditioned on $W > t$.}  We next identify the distribution of $Z = \Psi / W_t$.  The CDF of $Z$ conditioned on $W > t$ is denoted $F_{Z|t}(z)$, note this does not mean the distribution of $Z$ conditioned on $Z > t$.
\begin{eqnarray}
F_{Z|t}(z) &=& \int_t^{\infty} F_{\Psi}(zw) {\rm d}F_{W|t}(w) \nonumber \\
&=& \frac{1}{\bar{F}_W(t)} \int_t^{\infty} F_{\Psi}(zw) {\rm d}F_W(w). \label{app:am}
\end{eqnarray}
The PDF is
\begin{equation}
\label{app:an}
f_{Z|t}(z) = \frac{1}{\bar{F}_W(t)} \int_t^{\infty} w f_{\Psi}(zw) {\rm d}F_W(w).
\end{equation}

{\bf Step 3: Fractional order moment of $Z$.}  We next develop an expression for $\mathbb{E}[Z^{\delta}]$:
\begin{eqnarray}
\mathbb{E}[Z^{\delta}] &=& \int_0^{\infty} z^{\delta} {\rm d}F_{Z|t}(z) \nonumber \\
&=& \int_0^{\infty} z^{\delta} \left[ \frac{1}{\bar{F}_W(t)} \int_t^{\infty} w f_{\Psi}(zw) {\rm d}F_W(w) \right] {\rm d}z \nonumber \\
&=& \frac{1}{\bar{F}_W(t)} \int_t^{\infty} w \left[ \int_0^{\infty} z^{\delta} f_{\Psi}(zw) dz \right] {\rm d}F_W(w) \nonumber \\
&=& \frac{1}{\bar{F}_W(t)} \int_t^{\infty} w \left[ \int_0^{\infty} \left( \frac{x}{w} \right)^{\delta} f_{\Psi}(x) \frac{1}{w} {\rm d}x \right] {\rm d}F_W(w) \nonumber \\
&=& \frac{\mathbb{E}[\Psi^{\delta}]}{\bar{F}_W(t)} \int_t^{\infty} w^{-\delta}  {\rm d}F_W(w) \\
&=& \mathbb{E}[\Psi^{\delta}]\frac{\mathbb{E}[W^{-\delta} \mathbf{1}_{W > t}]}{\bar{F}_W(t)}. \label{app:ao}
\end{eqnarray}
This last expression gives the fractional moment in terms of $F_{\Psi},F_W$ given in the Theorem.  The distribution of $W$ may be used to obtain the expression in terms of the distributions $F_{\Psi},F_D$:
\begin{eqnarray}
\mathbb{E}[Z^{\delta}]
&=& \frac{\mathbb{E}[\Psi^{\delta}]}{\bar{F}_W(t)} \int_t^{\infty} w^{-\delta}  \left[ \int_0^{\infty} x^{\alpha} f_{\Psi}(w x^{\alpha}) {\rm d}F_D(x) \right] {\rm d}w \nonumber \\
&=& \frac{\mathbb{E}[\Psi^{\delta}]}{\bar{F}_W(t)} \int_0^{\infty} x^{\alpha} \left[ \int_t^{\infty} w^{-\delta} f_{\Psi}(w x^{\alpha}) {\rm d}w \right] {\rm d}F_D(x) \nonumber \\
&=& \frac{\mathbb{E}[\Psi^{\delta}]}{\bar{F}_W(t)} \int_0^{\infty} x^{\alpha} \left[ \int_{t x^{\alpha}}^{\infty} v^{-\delta} x^{2-\alpha} {\rm d}F_{\Psi}(v) \right] {\rm d}F_D(x) \nonumber \\
&=& \frac{\mathbb{E}[\Psi^{\delta}]}{\bar{F}_W(t)} \int_0^{\infty} x^2 \left[ \int_{t x^{\alpha}}^{\infty} v^{-\delta} {\rm d}F_{\Psi}(v) \right] {\rm d}F_D(x) \nonumber \\
&=& \mathbb{E}[\Psi^{\delta}] \frac{\int_0^{\infty} x^2 \left[ \int_{t x^{\alpha}}^{\infty} v^{-\delta} {\rm d}F_{\Psi}(v) \right] {\rm d}F_D(x)}{\int_0^{\infty} \left[ \int_{t x^{\alpha}}^{\infty} {\rm d}F_{\Psi}(v) \right] {\rm d}F_D(x)}. \label{app:ap}
\end{eqnarray}
The above development is elementary, involving only exchanging the order of integration and introducing a change of variables. $\blacksquare$

%%%%%%%%%%%%%%%%%%%%%%%%%%%%%%%%%%%%%%%%%%%%%%%%%%%%%%%%%%%%%%%%%%%%%%%
% A-5. Proof of Theorem 5
%%%%%%%%%%%%%%%%%%%%%%%%%%%%%%%%%%%%%%%%%%%%%%%%%%%%%%%%%%%%%%%%%%%%%%%
\section*{Proof of Theorem \ref{thm:5}}

Theorem \ref{thm:5} asserts that the sole effect of changing from randomized transmission decisions (with channel inversion) to threshold based transmission decisions (with channel inversion) is to change the distribution of the received interference power.  Consider the expressions for the normalized aggregate interference seen by the reference receiver, $Y$, under randomized transmissions (\ref{eqn:aa}) and under threshold transmissions (\ref{eqn:bl}).  They are the same with the exception that $W_i$ in (\ref{eqn:aa}) is replaced with $W_{i|t}$ in (\ref{eqn:bl}).  Given that all the results in Corollaries \ref{cor:1}, \ref{cor:2}, and \ref{cor:3} depend upon the fractional order moment $\mathbb{E}[Z^{\delta}]$, for $Z = \Psi/W$, it follows that these same results will hold under threshold transmissions, with $Z_t = \Psi/W_t$, and fractional order moment $\mathbb{E}[Z_t^{\delta}]$.  All that remains is to establish the bounds on the transmission capacity given by (\ref{eqn:bn}).

Let $\theta$ be the constant given by (\ref{eqn:ae}), and let $\mu \in [0,\lambda]$ be a generic intensity of attempted transmissions.  Define $g_{\rm max} = \theta \lambda$.  The outage probability bounds in (\ref{eqn:ad}) depend upon $\mu,\theta$ only through the product $g = \theta \mu$; think of $\epsilon^l = q^l(g)$, $\epsilon^u = q^u(g)$ in (\ref{eqn:ad}) as the lower and upper outage probabilities for a normalized intensity of transmission attempts $g \in [0,g_{\rm max}]$.  The bounds are both bijections on $[0,1]$, as such they admit unique inverses, denoted $g^l= q^{l,-1}(\epsilon)$ and $g^u = q^{u,-1}(\epsilon)$.  Think of $g^l,g^u$ as the bounds on the normalized transmission attempt intensity required for outage probability $\epsilon$.

Under the threshold decision rule both $\mu,\theta$ depend upon $t$.  The function $\gamma(t) = \theta(t) \mu(t)$ gives the normalized transmission attempt intensity under each threshold $t \in \mathbb{R}^+$.  It is clear that $\gamma(t)$ is monotone decreasing in $t$ onto the interval $[0,g_{\rm max}]$, and as such it too admits a unique inverse, denoted $\gamma^{-1}(g)$ for $g \in [0,g_{\rm max}]$.  Think of $t = \gamma^{-1}(g)$ as the threshold required for a normalized transmission attempt intensity of $g$.  To summarize, each threshold $t \in \mathbb{R}^+$ maps to a normalized intensity of attempted transmissions $g = \gamma(t)$, and each $g$ maps to bounds on the outage probability given by $\epsilon^l = q^l(g), \epsilon^u = q^u(g)$, for $q^l(g), q^u(g)$ given in (\ref{eqn:ad}).

The optimal contention density, $\nu(\epsilon)$, has an associated optimal threshold, $t(\epsilon)$, such that $\nu(\epsilon) = \lambda \bar{F}_W(t(\epsilon))$, where $W = \Psi D^{-\alpha}$.  That is, $\nu(\epsilon)$ is the maximum intensity of transmission attempts with an associated outage probability of $\epsilon$, but this may also be expressed as the intensity of potential transmitters, $\lambda$, thinned by the probability that a typical potential transmitter's signal strength, $W$, exceeds some threshold $t(\epsilon)$.   Using the above definitions,
\begin{equation}
t^l(\epsilon) = \gamma^{-1}(q^{l,-1}(\epsilon)), ~~
t^u(\epsilon) = \gamma^{-1}(q^{u,-1}(\epsilon)),
\end{equation}
with associated bounds on the optimal contention density given by
\begin{equation}
\nu^l(\epsilon) = \lambda \bar{F}_W(t^u(\epsilon)), ~~
\nu^u(\epsilon) = \lambda \bar{F}_W(t^l(\epsilon)).
\end{equation}
Finally, the transmission capacity bounds are simply
\begin{equation}
c^l(\epsilon) = \nu^l(\epsilon) (1-\epsilon), ~~
c^u(\epsilon) = \nu^u(\epsilon) (1-\epsilon).
\end{equation}
The requirement that $g \in [0,g_{\rm max}]$ translates to an upper bound on $\epsilon$ such that $\epsilon$ in $c^l(\epsilon)$ must satisfy $\epsilon \leq q^u(g_{\rm max})$, and $\epsilon$ in $c^u(\epsilon)$ must satisfy $\epsilon \leq q^l(g_{\rm max})$.  $\blacksquare$

\clearpage
\onecolumn

\begin{figure}[ht]
\centering
\includegraphics[width=3.5in]{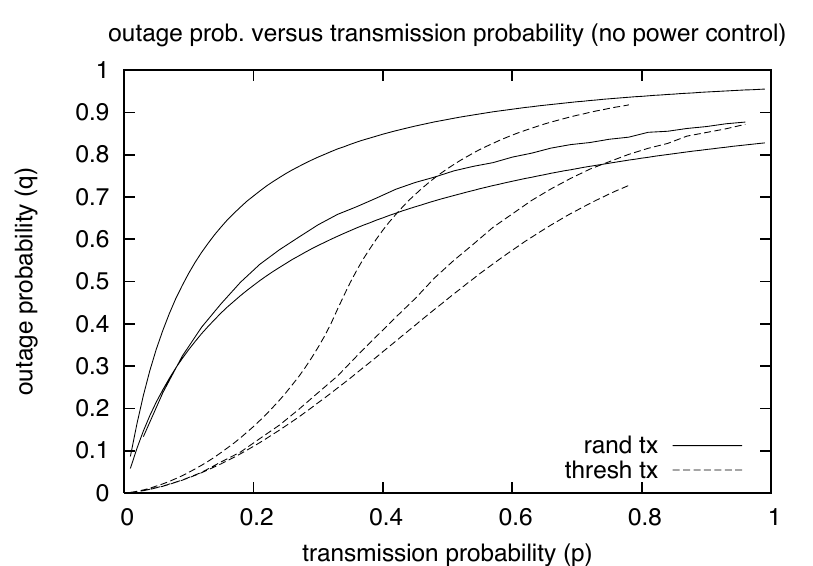}
\includegraphics[width=3.5in]{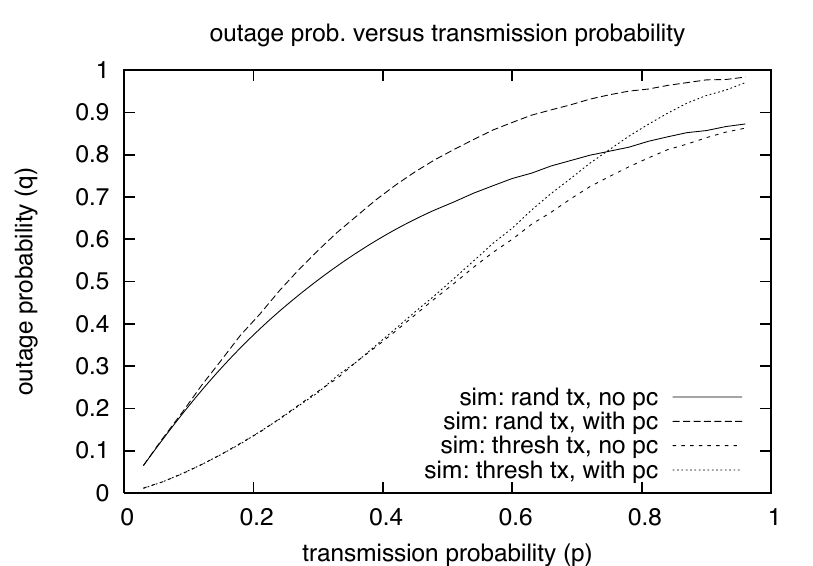}
\includegraphics[width=3.5in]{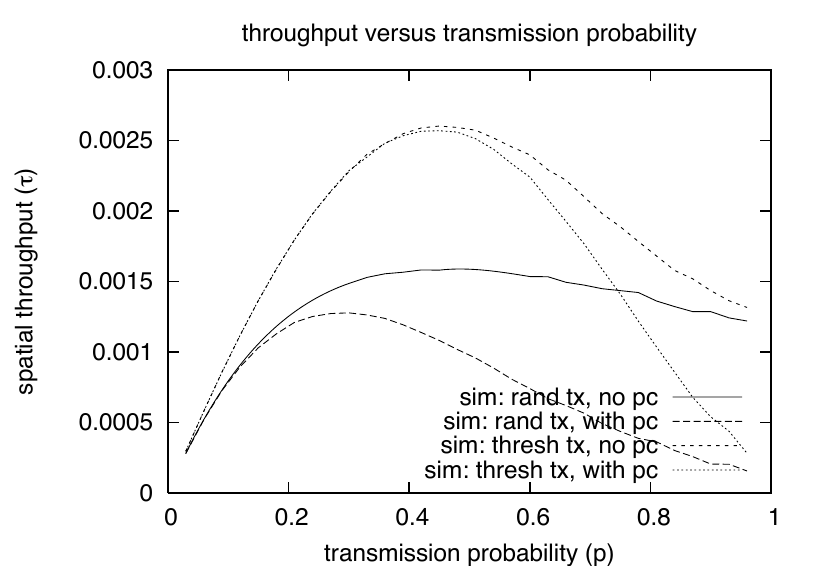}
\includegraphics[width=3.5in]{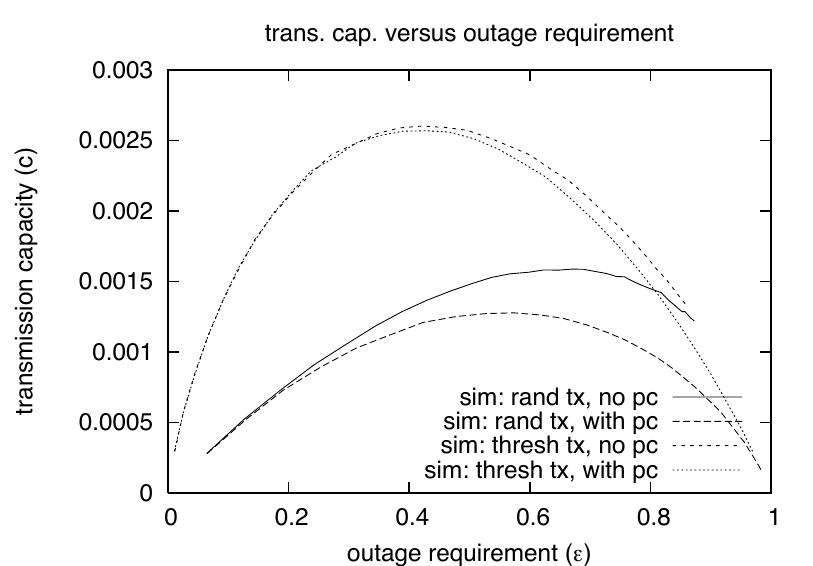}
\caption{{\bf Example 1: Lognormal shadowing.}  All plots are for $\sigma =  \frac{\log 10}{10} 6$ ($6$ dB).  {\bf Top left:} outage probability $q$ versus the transmission probability $p$ for both randomized transmissions (solid curves) and threshold transmissions (dashed curves) with no channel inversion.  The three curves for each case are the lower and upper bounds along with simulation results.  {\bf Other plots:} The other three plots show simulation results for the four cases: randomized transmissions with and without channel inversion, and threshold transmissions with and without channel inversion.}
\label{fig:1}
\end{figure}

\begin{figure}[ht]
\centering
\includegraphics[width=3.5in]{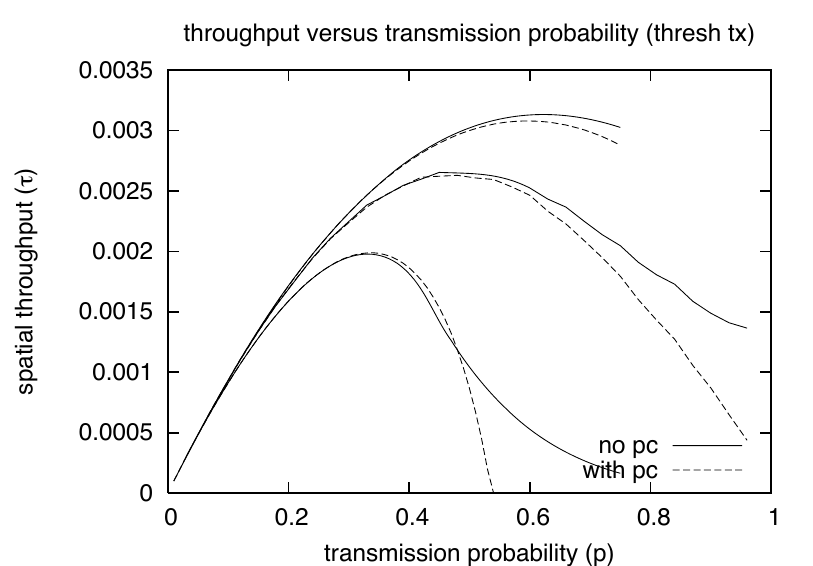}
\includegraphics[width=3.5in]{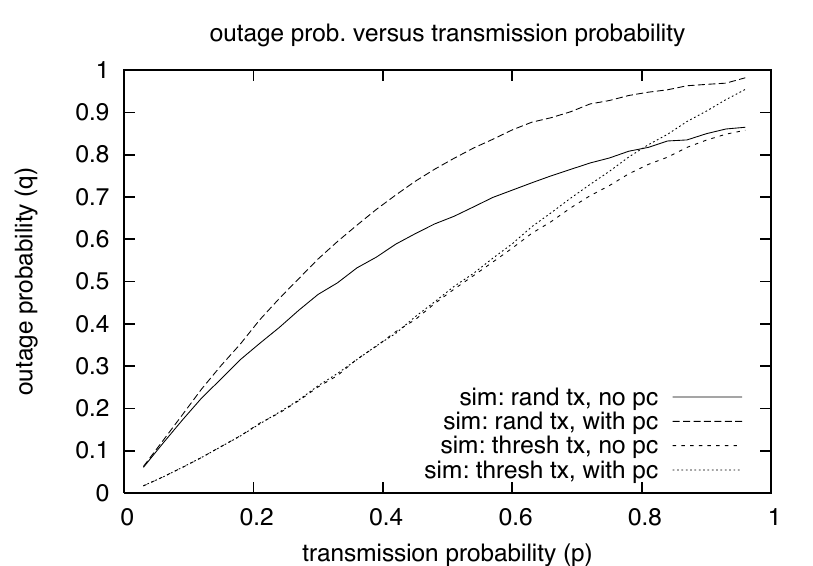}
\includegraphics[width=3.5in]{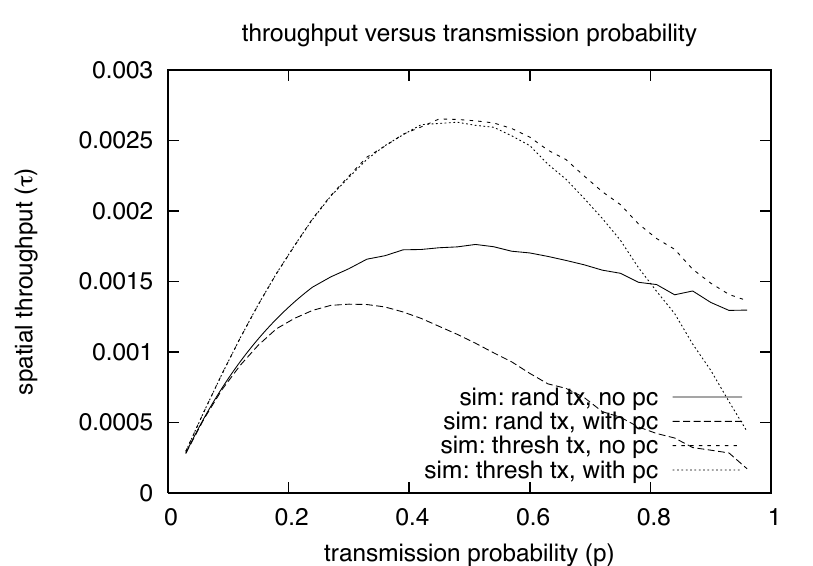}
\includegraphics[width=3.5in]{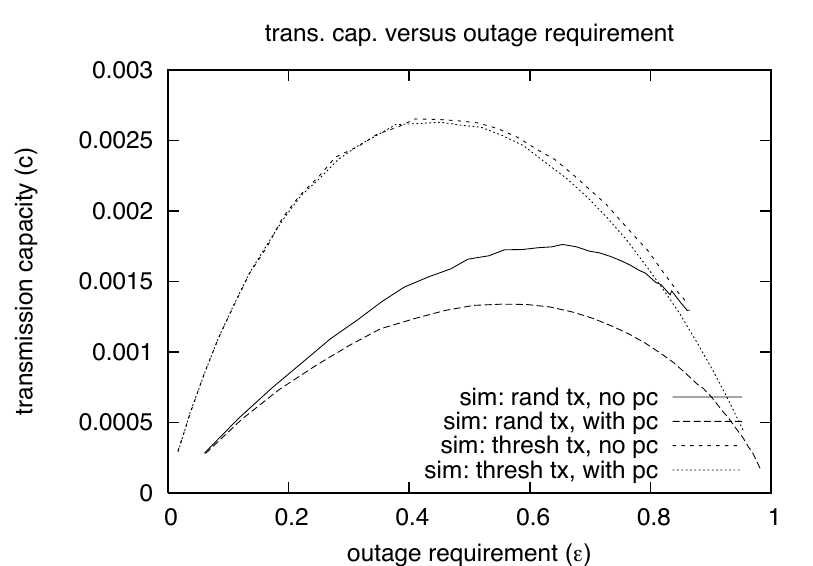}
\caption{{\bf Example 2: Rayleigh fading.}  {\bf Top left:} Spatial throughput $\tau$ versus the transmission probability $p$ for threshold based transmissions with channel inversion (dashed curves) and without channel inversion (solid curves).  The three curves for each case are the lower and upper bounds along with simulation results.  {\bf Other plots:} The other three plots show
simulation results for the four cases: randomized transmissions with and without channel inversion, and threshold transmissions with and without channel inversion.}
\label{fig:2}
\end{figure}

\begin{figure}[ht]
\centering
\includegraphics[width=3.5in]{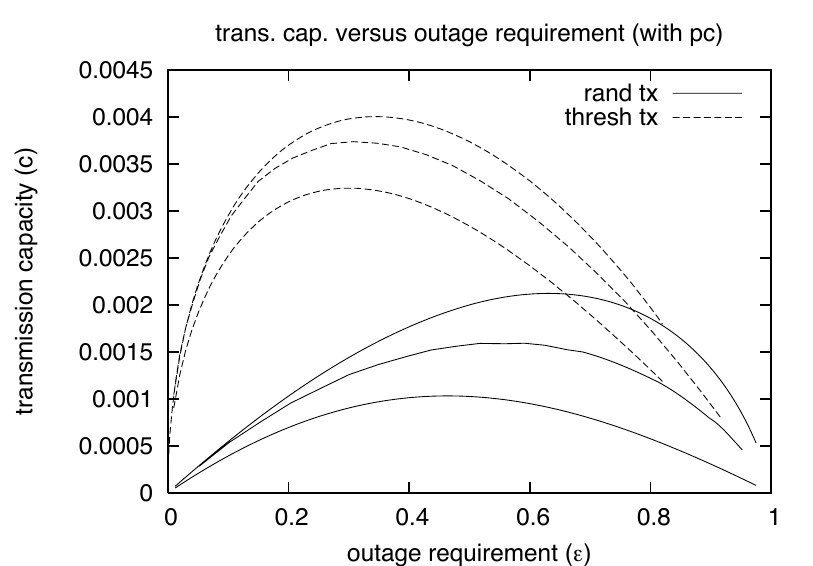}
\includegraphics[width=3.5in]{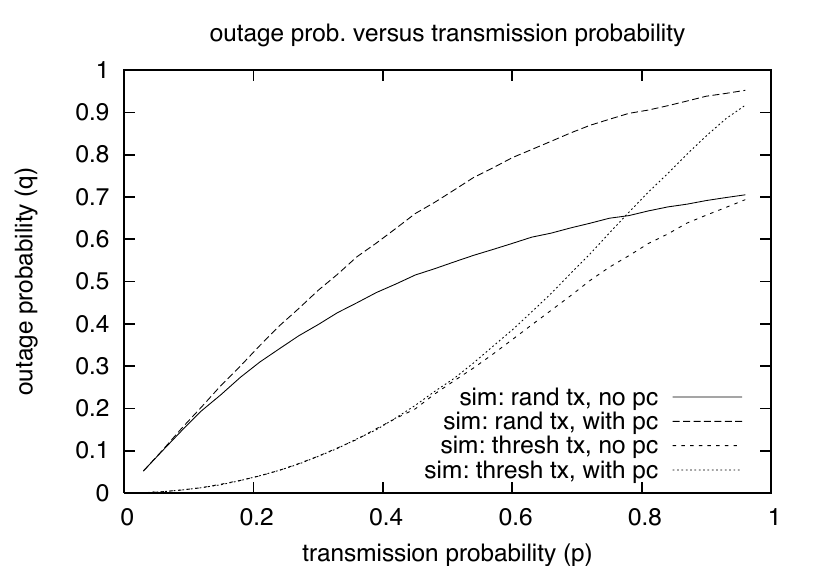}
\includegraphics[width=3.5in]{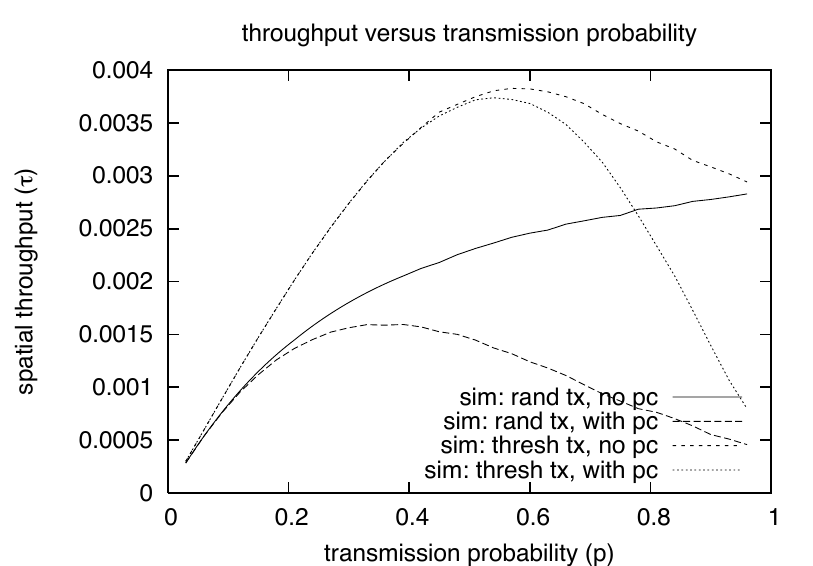}
\includegraphics[width=3.5in]{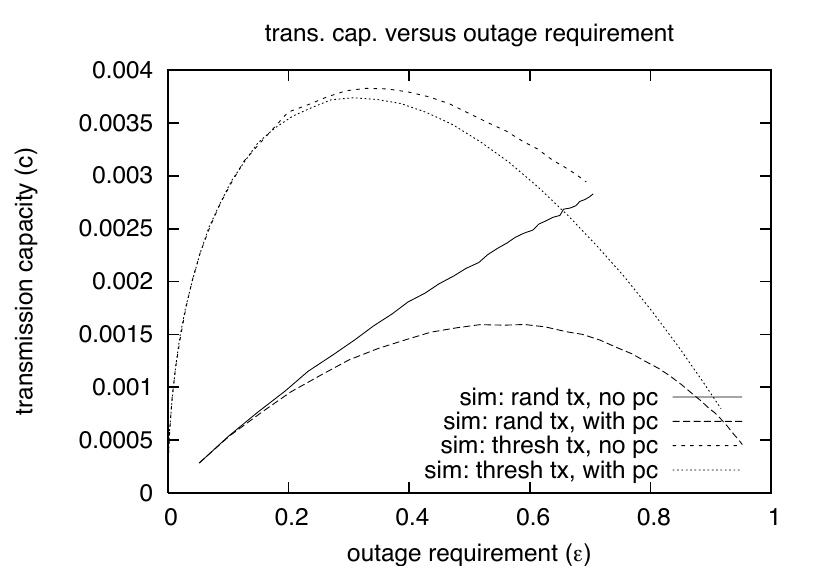}
\caption{{\bf Example 3: Nearest receiver transmissions.}  {\bf Top left:} Transmission capacity $c$ versus the outage requirement $\epsilon$ for randomized transmissions (solid curves) and threshold transmissions (dashed curves) with channel inversion.  The three curves for each case are the lower and upper bounds along with simulation results. {\bf Other plots:} The other three plots show simulation results for the four cases: randomized transmissions with and without channel inversion, and threshold transmissions with and without channel inversion.}
\label{fig:3}
\end{figure}

\end{document}